\documentclass[12pt]{iopart}
\usepackage{graphicx,color,epsfig,rotate}
\usepackage{amssymb}
\begin{document}
\title[Orbital and spin effects for the upper critical
field ]
{Orbital and spin effects for the upper critical
field in As deficient disordered Fe pnictide superconductors} 
\author{G~Fuchs$^1$, S-L~Drechsler$^1$, N~Kozlova$^1$, 
M~Bartkowiak$^2$, JE~Hamann-Borrero$^1$, 
G~Behr$^1$,
K~Nenkov$^1$, H-H~Klauss$^3$,
H~Maeter$^3$, A~Amato$^4$,   H~Luetkens$^4$, A~Kwadrin$^3$,
R~Khasanov$^4$,  J~Freudenberger$^1$,  A~K\"ohler$^1$,
M~Knupfer$^1$, E~Arushanov$^5$, H~Rosner$^6$,
B~B\"uchner$^1$, and L~Schultz$^1$}
\address{$^1$ Leibniz-Institut IFW Dresden, P.O. Box 270116, D-01171 
Dresden, Germany}
\address{$^2$ Leibniz-Institut FZ Dresden-Rossendorf (FZD), Germany}
\address{$^3$ Institut f\"ur Festk\"orperphysik, TU Dresden, 
Dresden, Germany}
\address{$^4$ Laboratory f.\ Muon-Spin Spectroscopy, Paul Scherrer
Institut, 
Villigen, 
Switzerland}
\address{$^5$ Institute of Applied Physics, Acad.\ of Sciences of 
Moldova, 
Chisinau,  Moldova }
\address{$^6$ Max-Planck-Institut f.\ Chemische Physik fester Stoffe, 
Dresden, Germany}
\ead{drechsler@ifw-dresden.de}

\begin{abstract}
We report   
upper critical field $B_{c2}(T)$ data for 
LaO$_{0.9}$F$_{0.1}$FeAs$_{1-\delta}$ in a wide temperature and 
field range up to 
60~T. The
large  
slope of $B_{c2} \approx $~-5.4 to -6.6T/K 
near an {\it improved} $T_c \approx $ 28.5~K 
of
the 
in-plane 
$B_{c2}(T)$ contrasts with a 
flattening starting 
near 23~K above 30~T 
we regard as the onset of  
Pauli-limited behavior (PLB) with $B_{c2}(0)\approx$~63 to 68~T. 
We interpret a similar hitherto unexplained flattening of the 
$B_{c2}(T)$ curves 
reported for at least three 
other disordered closely related systems as 
the Co-doped BaFe$_2$As$_2$, the (Ba,K)Fe$_2$As$_2$,
or the NdO$_{0.7}$F$_{0.3}$FeAs
 (all single crystals) for 
applied fields $H \parallel (a,b)$ also as a manifestation of PLB.
Their Maki parameters
have been estimated analyzing  their 
$B_{c2}(T)$ data within the Werthamer-Helfand-Hohenberg approach. 
The pronounced PLB of
(Ba,K)Fe$_2$As$_2$
 single crystals obtained from a Sn-flux
is attributed also
to a significant As deficiency 
detected by
wave length dispersive x-ray spectroscopy 
as reported by 
Ni~N {\it et al.} 2008 {\it Phys.\ Rev.\ }B {\bf 78} 014507.  
Consequences of our results are discussed 
in terms of disorder effects within
conventional (CSC) and unconventional 
superconductivity (USC). 
USC scenarios with nodes on individual Fermi surface sheets (FSS), e.g.\ 
$p$- and $d$-wave SC,
can be discarded for our samples.
The increase of d$B_{c2}$/d$T\mid_{T_c}$ 
by sizeable disorder
provides 
evidence for an important {\it intraband} (intra-FSS) contribution
to the orbital upper critical field.
We suggest that it can be ascribed either to an impurity driven transition
from 
$s_{\pm}$ USC to 
CSC of an
extended $s_{++}$-wave state or to a stabilized $s_{\pm}$-state provided
As-vacancies cause predominantly strong intraband scattering in the 
unitary limit.  
We compare
our results with $B_{c2}$ data from the literature which show often no 
PLB 
for fields below 60 to 70~T probed so far. 
A novel disorder related scenario of a complex 
interplay of 
SC with two different competing magnetic instabilities
is suggested.

\end{abstract}
\submitto{\NJP}
\maketitle
\normalsize

\section*{\bf Content}

{\textcolor{blue}{
\bf 1. \hspace{0.6cm}Introduction \hfill 2\\
2.\hspace{0.70cm}Experiment \hfill 4\\
2.1\quad \hspace{0.2cm} Sample preparation and lattice constants \hfill 4\\
2.2\quad \hspace{0.2cm} Measurements and used experimental techniques \hfill 6\\
3.\hspace{0.73cm}Results \hfill 6\\
3.1 \quad	Resistivity data \hfill 6\\
3.2 \quad Enhanced paramagnetism: Muon spin rotation measurements \hfill 8\\
4. \hspace{0.58cm}  Upper critical field \hfill  9\\
4.1 \quad	Resistance for applied static and pulsed fields \hfill 9\\ 
4.2 \quad	Different criteria for the determination of  $B_{c2}$ \hfill 11\\
5.\hspace{0.70cm}Analysis of $B_{c2}(T)$ and discussion \hfill 13\\
5.1 \quad Orbital and paramagnetic upper critical field \hfill 13\\
5.2 \quad As deficient LaO$_{0.9}$F$_{0.1}$FeAs$_{1-\delta}$ \hfill 15\\
5.3 \quad Comparison with other samples: slope of $B_{c2}(T)$ near $T_c$, 
disorder, \\
\qquad \ \ and paramagnetism \hfill 15\\
5.4.\quad Aspects of anisotropy and multiband superconductivity \hfill 18\\
5.5 \quad Possible origin of the Pauli-limiting behaviour \hfill 20\\
6.\hspace{0.69cm}Conclusions and Outlook \hfill 22\\
Acknowledgments \hfill 24\\
References \hfill 24\\
Supplementary material \hfill 28\\
S1 \quad	Hall data \hfill 28\\
S2 \quad 	Scaling analysis of the resistivity \hfill 28\\
}\\

\section{\bf Introduction}

The recently achieved relatively high superconducting transition
temperatures $T_c$ up to 57 K and last but not least remarkably 
high upper critical 
fields $B_{c2}(0)$ exceeding often at least 70 T of Fe pnictides 
followed their discovery in the system 
LaO$_{1-x}$F$_{x}$FeAs \cite{kamihara08,cheng} 
has opened the door to a fascinating 
world of novel superconductors. 
Naturally, shortly after the discovery of these novel FeAs based 
superconductors,   
the underlying 
pairing mechanism and many basic physical properties both in the
superconducting and in the normal state are still not well understood.
In this context one can only agree with the 
statement "There is a serious need to identify and address relatively 
straightforward questions, in addition to broader investigations to compare
and contrast " all the Fe pnictides and related materials " to identify 
trends that might provide a clue" \cite{lee-alphaFeSe}.
Such general questions are:
(i) Is this superconductivity (SC) based on Cooper pairs or on bipolarons?
(ii) What is  the symmetry of its order parameter ?
(iii) How is the SC affected by disorder?
A study of the upper critical field $B_{c2}(T)$ 
as a fundamental quantity of the SC is expected to
provide valuable insight into the nature of the interaction responsible
for the formation of Cooper pairs and to help us
to answer questions (i)-(iii) in near future.
Since
the usual $el$-$ph$ mechanism has been ruled out by a much too
weak coupling strength $\lambda \leq$~0.2 \cite{boeri}, 
a variety of nonstandard mechanisms mostly involving spin fluctuations 
has been proposed \cite{mazin1,korshunov,chubukov}. 
This way it might also provide  constraints
for 
proposed unconventional scenarios \cite{symrem}
based on repulsive interactions.
Concerning the symmetry of the superconducting order parameter
we note its robustness or sensitivity to various scattering processes.  
In this context the so-called sign reversal isotropic \cite{node} $s_{\pm}$ interband 
scenario
\cite{mazin1,chubukov,mazin-schmalian,dolgov1,dolgov09,ummarino,senga1, senga2} 
is of special interest.
Here a repulsive interband interaction between disconnected
nearly nested hole-type ($h$) and electron-type ($el$) Fermi surface 
sheets (FSS) 
is suggested to be nearly as effective in creating superconductivity
as a standard attractive one. This, as predominantly assumed 
magnetic interaction is thought to be
responsible for opposite signs of the  
superconducting order
parameter (gap) on $h$- and $el$
FSS centred around the $\Gamma$-point 
and the corners of the Brillouin zone, respectively.
The resonance peak observed recently below $T_c\approx 38$~K near a 
transferred energy of 
14~meV
and transferred momentum of $Q=1.15$\AA$^{-1}$ 
in recent inelastic neutron scattering measurements on the 
122 system Ba$_{0.6}$K$_{0.4}$Fe$_{2}$As$_{2}$ \cite{christianson}
has been regarded as evidence for an unconventional $s_{\pm}$-pairing
state.
 However, the strength of that interband coupling compared with the intraband 
interactions and its related stability against a competing 
conventional $s_{++}$-pairing 
triggered by an enhanced $s_{\pm}$-pairbreaking 
due to a possibly enlarged
interband nonmagnetic impurity 
scattering and/or a reduced interband coupling 
due to a smeared nesting
remains unclear \cite{rems}. More sophisticated studies along 
these lines are necessary to settle these questions.

In such a confusing situation a combination of several approaches
seems to be necessary: (i)
a detailed
study of selected well-defined systems with  a controlled amount
of deviations of stoichiometry and other kinds of disorder,
(ii) a systematic
comparison of various members of the fastly growing 
FeAs family 
with the so far discovered three classes of 1111, 122, and 111 systems
(see the other contributions of this volume and below)
including also ferroselenides, -tellurides, -phosphides and
other related layered compounds
(iii) a comparison also with other exotic superconductors.

In general, the area of very high magnetic fields is experimentally
not well studied due to the large necessary technical efforts 
 and restricted mainly to resistance measurements
for pulsed field magnets where the highest fields exceeding 40~T
are achieved so far \cite{high-field}. For that reason fields up
to 60~T as reported here are available in few 
laboratories 
worldwide, only. For completeness it should be noticed that
theoretically a rich variety of unexpected
phenomena has been predicted \cite{rasolt} for ultra-high 
magnetic fields
being a challenge for further future studies.

Controlled disorder provides insight into relevant scattering 
processes and in the symmetry of the pairing since very often an
unconventional pairing in the 
sense of $T_c$ and $dB_{c2}/dT |_{T_c}$ \cite{posazhennikova,charikova}
is expected to be more or less strongly suppressed by 
disorder. 
Adopting for instance 
the so-called self-consistent Born approximation
valid for relatively weak scattering, one has for $T_c$ 
 \cite{lin,petrovic,mackenzie,radtke}:
\begin{equation}
-\ln \left( \frac{T_c}{T_{c0}} \right)=\psi \left( \frac{1}{2}+
\frac{\beta T_{c0}}{2\pi T_c} \right) -\psi \left( \frac{1}{2}\right), 
\end{equation}
where $\psi (x)$ is the digamma function and $\beta$ is the 
 strong-coupling 
pair-breaking parameter
$\beta = \Omega^2_p \rho_0/8\pi (1+\lambda) T_{c0}$ 
which is 
related to the residual resistivity $\rho_0$ and the plasma energy
$\Omega_p$ in the $(a,b)$-plane. 
However, it should be noted that the $T_c$-suppression
in the opposite limit of strong scattering (unitary limit) 
is less prononounced. In particular,
it has been suggested that in this unitary limit 
for a two-band superconductor in the unconventional $s_{\pm}$-regime, 
a weaker pair-breaking interband scattering (compared 
with non-pair-breaking intraband one) will
practically drop out \cite{senga1,senga2}. Anyhow, the relevance 
of this approach
to the As-vacancy case considered here remains unclear 
and further theoretical
studies of the scattering properties of As vacancies as well as for other
impurities such as Co and Ni on Fe-sites
are highly desirable.
Some, 
but much weaker, suppression might occur
also in the anisotropic or multiband conventional $s_{++}$-wave case
since the scattering may smear out the gap 
anisotropy. However, it will be shown that surprisingly 
just the opposite, namely, an enhancement of $T_c$
happens in our case.

For low applied fields rather different slopes  
$dB_{c2}/dT\approx$~-1.6 T/K up to -2 T/K at $T_c\approx$~26~K
 \cite{sefat,hunte} and up to -4~T/K at $T_c\approx$ 20~K~ \cite{chen1}
have been reported
for the As stoichiometric La based compounds. Here, 
we report with $dB_{c2}/dT\approx$~-5.4 to -6.6 T/K, 
to our knowledge one of the highest 
slopes of $B_{c2}$ near $T_c$ observed so far
for the La-series. Another interesting issue of high-field studies
considered here is the possibility to observe Pauli-limiting behavior (PLB).
Triplet $p$-wave pairing 
or  strong coupling ($B_{c2}(0)\geq$ 60~T) 
would naturally explain the reported absence of PLB \cite{hunte}. 
In this context
it is important to note that we succeeded in detecting PLB for our specific
sample. It points to $B_{c2}(0)$-values
being much below often used 
WHH (Werthamer-Helfand-Hohenberg) \cite{WHH} 
based estimates.
After presenting 
various data 
 which deviate 
from those of Ref.~\cite{hunte} as well as 
from 
our As-non-deficient quasi-clean samples 
\cite{luetkens,klauss,drechsler,graefe,hess}, we will discuss our $B_{c2}(T)$
data in the light of these more general issues.

\section{\bf Experiment}
\subsection{Sample preparation and lattice constants}
\begin{figure}[t] 
\hspace{-0.3cm} 
\includegraphics[width=8cm]{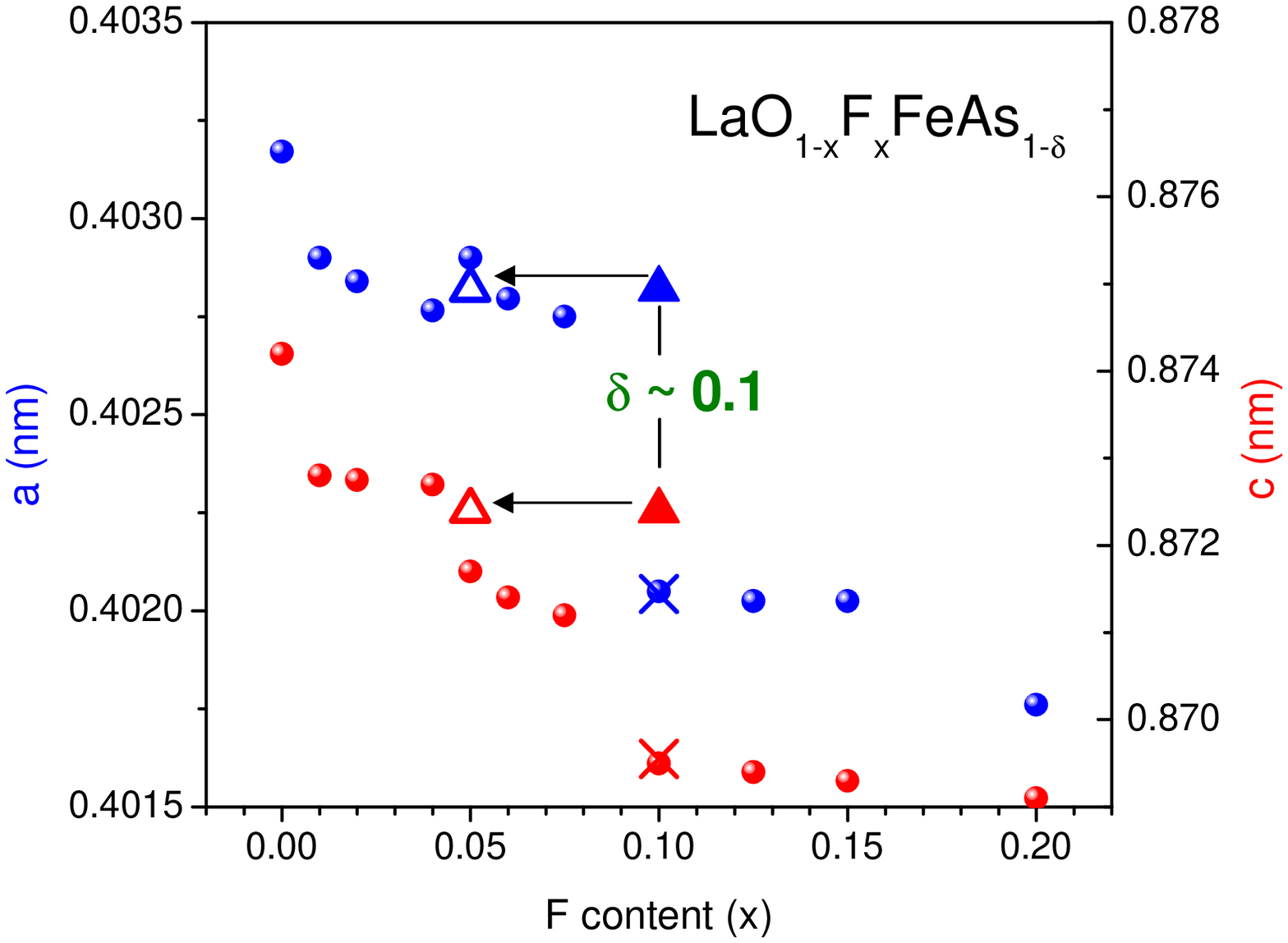}
\includegraphics[width=7.2cm]{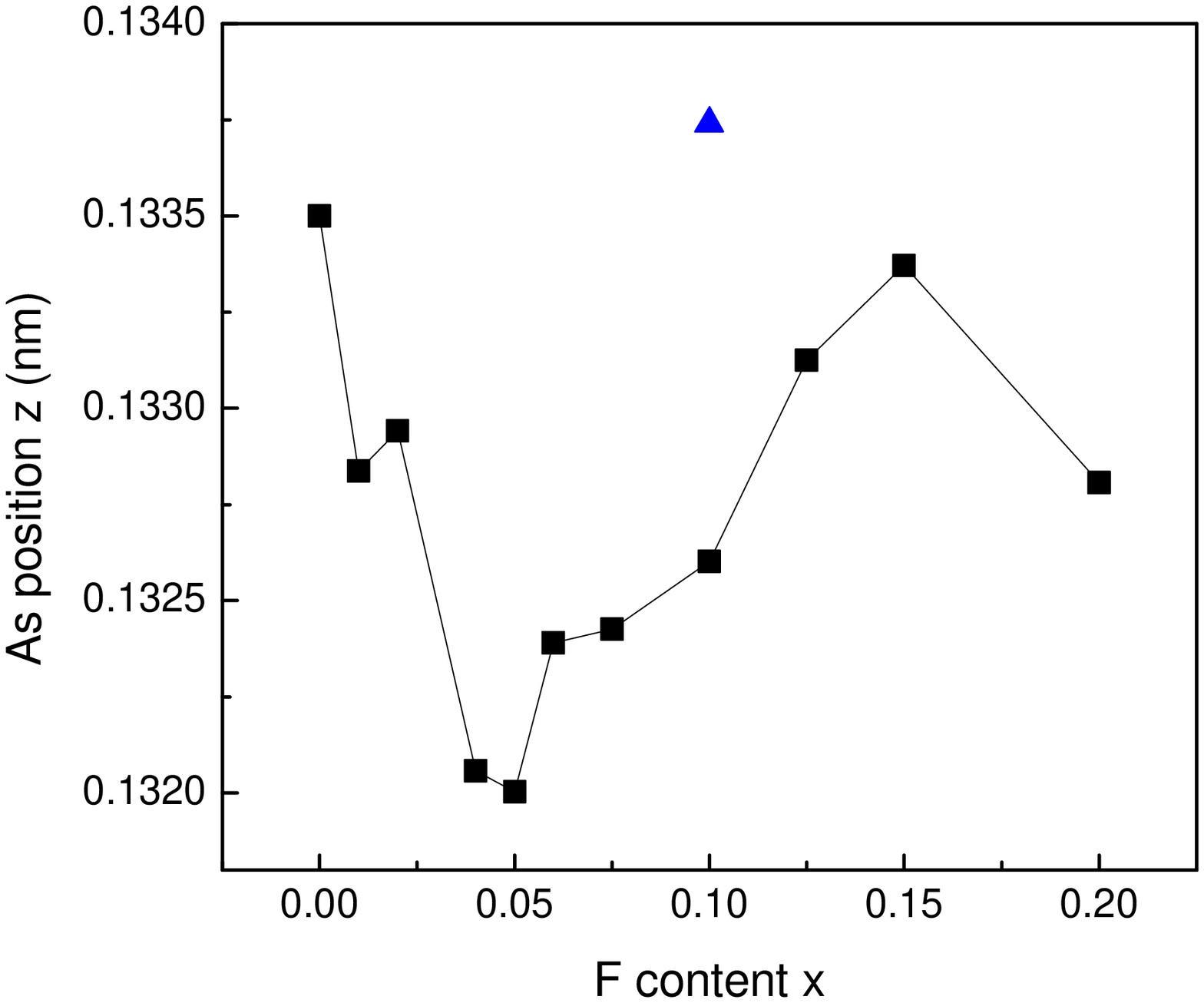}
\hspace{0.7cm}
\caption{Left: lattice constants of 
various LaO$_{1-x}$F$_x$FeAs samples for different 
F-content
(from undoped $x=0$ (left) to the overdoped case $x=0.2$ 
(right);
\textcolor{red}{$\bullet$} - $c$-axis data, 
\textcolor{blue}{$\bullet$} - $a$-axis data taken from reference
\cite{luetkens}, $\textcolor{red}{\times}$, 
$\textcolor{blue}{\times}$ - clean reference sample ($x=0.1$), 
$\textcolor{red}{\blacktriangle}$, $\textcolor{blue}{\blacktriangle} $ - 
As-deficient sample; $\textcolor{red}{\triangle}$, 
$\textcolor{blue}{\triangle} $ - nearly equivalent non-deficient sample
with  $x \approx 0.05$ for 
comparison. Right: As-positions in the non-deficient 
($\blacksquare$)
and in the deficient ($\textcolor{blue}{\blacktriangle}$) samples. } 
\label{f1}
\end{figure}
Polycrystalline samples of LaF$_{0.1}$O$_{0.9}$FeAs 
were prepared from pure components (3$N$ or better) 
using a two-step solid state reaction method similar to that described 
by Zhu {\it et al.} \cite{zhu-hall}. In the first step, Fe powder and
 powdered As particles were milled, mixed and pressed into pellets 
under Ar atmosphere, and annealed at 500~$^{\circ}$C for 2~h and at 
700~$^{\circ}$C for 10~h 
in an evacuated silica tube. In the second step, the Fe-As pellets were 
milled and mixed with La powder, annealed La$_2$O$_3$ powder, and anhydrous 
LaF$_3$ powder and subsequently pressed into pellets under a well-defined 
pressure. Then, the samples were heated in an evacuated silica tube at 
940$^{\circ}$C 
for 2~h and at 1150~$^{\circ}$C for 48~h. To improve the homogeneity, 
the 940~$^{\circ}$C annealing step was prolonged. Some samples 
have been  wrapped in a Ta~foil during the annealing procedure of the 
second step (see also reference \cite{fuchsprl}). Ta acts as an As 
getter at high temperatures forming a solid solution of about 9.5~at.{\%}~As 
in Ta 
with a small layer of Ta$_2$As and TaAs on top of the foil. This leads to an As 
loss in the pellets. The annealed pellets were ground and polished, and the 
local 
composition of the resulting samples was investigated by wavelength-dispersive 
and energy-dispersive x-ray spectroscopy (WDX and EDX, respectively) in a 
scanning electron microscope (SEM). The amount of impurity phases does not 
exceed the x-ray diffraction resolution limit of $\sim 5$~\%. 
According to the EDX analysis, an As/Fe ratio 
of about 1.0
was found in the reference sample annealed without a Ta~foil to be compared with
0.90 to 0.95 in the As-deficient sample.  
A powder-x-ray diffraction study with a Rietveld refinement of the 
main phase 
yields enhanced lattice constants of 
$a = 0.4028$~nm and 
$c = 0.8724$~nm for 
the As deficient sample compared to 
$a = 0.402$~nm and 
$c = 0.8696$~nm for the reference 
sample \cite{fuchsprl}. 
In figure~\ref{f1} (left), the lattice parameters of these two 
samples are included 
in the dependence of the lattice parameters on the 
nominal F content found for 
non-As-deficient LaF$_x$O$_{1-x}$FeAs samples studied at 
the IFW Dresden \cite{luetkens}.
We note that for $x > 0.04$ the nominal fluorine concentration practically 
coincides with that determined from the WDX-analysis. 
A continuous decrease of the lattice parameters with increasing F content is 
observed consistently with other reports. Whereas the lattice parameters of our 
reference sample well agree with the data for $x = 0.1$ shown in 
figure~1 (left), the 
lattice constants of the As deficient sample are close to those for underdoped samples 
near the border of magnetism and superconductivity at $x = 0.05$ for 
stoichiometric samples. 
The reduced charges of the anionic As  and that of 
cationic Fe layers causes less attraction between them and  
is thereby
responsible for the increase of the As-position (see figure 1 (right)).
According to an analysis of the reflectivity a similar amount of charge
gives rise to an additional optical absorption we ascribed to bound
electrons localised in the Fe-plane near the As-vacancies \cite{drechsler09}.
This explains why our As-deficient sample is not strongly overdoped by electrons
as one might expect at first glance and what is the microscopic reason
for the strongly enhanced scattering of the quasiparticles bearing the transport
and the superconductivity. The same electrostatic argument 
 explains a slight flattening
of the LaO$_{0.9}$F$_{0.1}$-bilayer as compared with the more elongated As-Fe$_2$-As trilayer
and the resulting increase of $c$.

\subsection{Measurements and used experimental techniques}

The electrical resistance and the Hall effect were measured for plate-like 
samples using the standard four-point method. These measurements were done 
in a Physical Property Measurement System (PPMS, Quantum Design) in fields 
up to 14~T. In addition,  resistance measurements were performed in the 
pulsed-field facilities of the IFW Dresden and the FZD up to 50~T 
and 60~T, respectively. Gold contacts (100~nm thick) were made by sputtering 
in order to provide a low contact resistivity and to avoid possible heating 
effects in the pulsed field measurements. Furthermore, some magnetic 
properties of 
our samples were 
studied
by muon spin relaxation
($\mu$SR) 
measurements which were performed at the Paul-Scherrer-Institute, Villigen, 
at zero field and in transverse applied fields both in the 
superconducting state at 1.6~K and in the normal state at 40~K.

\section{\bf Results}
\subsection{Resistivity data}    

In figure~2, the resistivity data for the clean reference and the 
As-deficient sample 
are compared. The resistivity 
\begin{figure}[t]
\hspace{0cm}
\includegraphics[width=7.5cm]{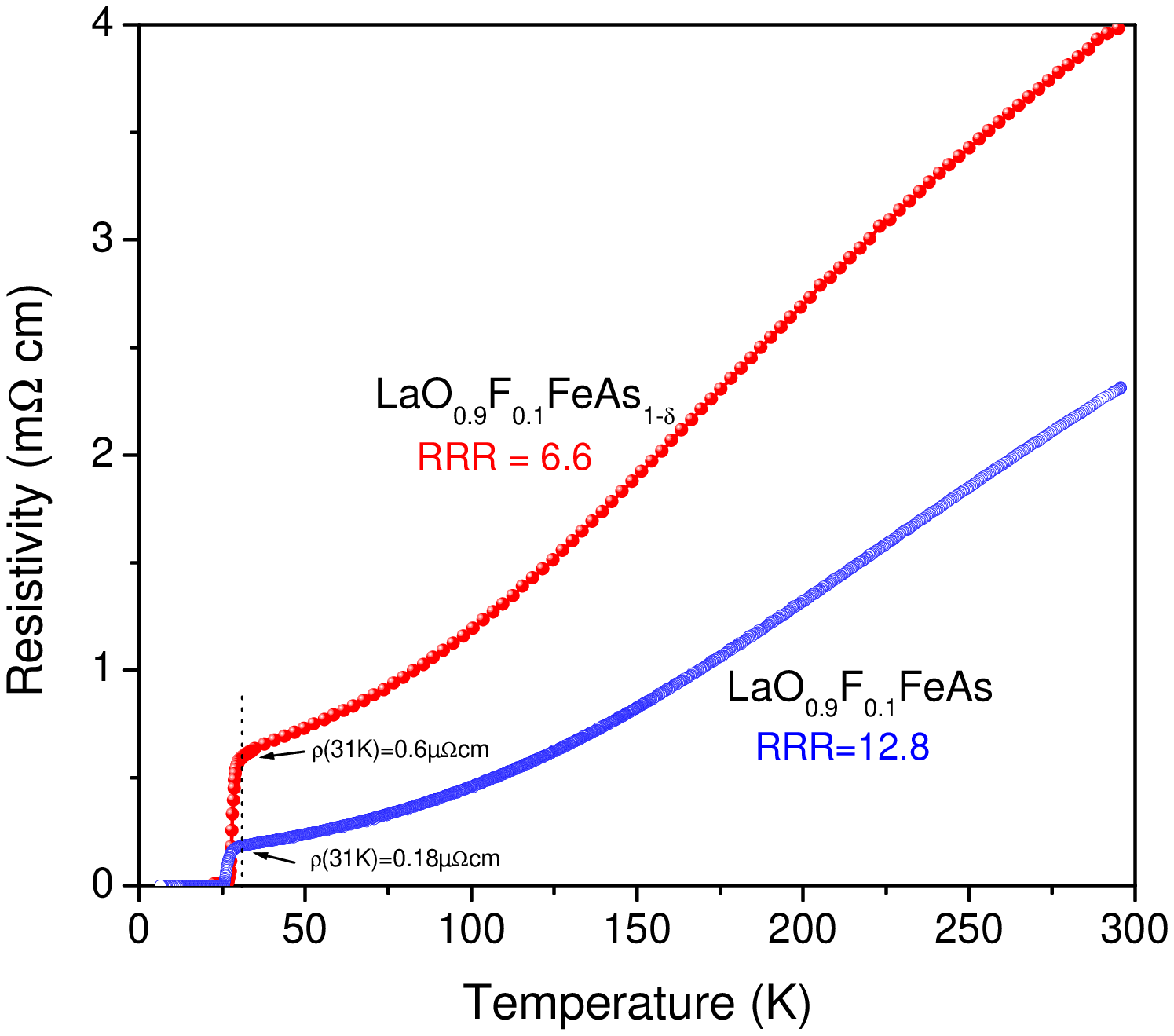}
\hspace{0.3cm}
\includegraphics[width=7.7cm]{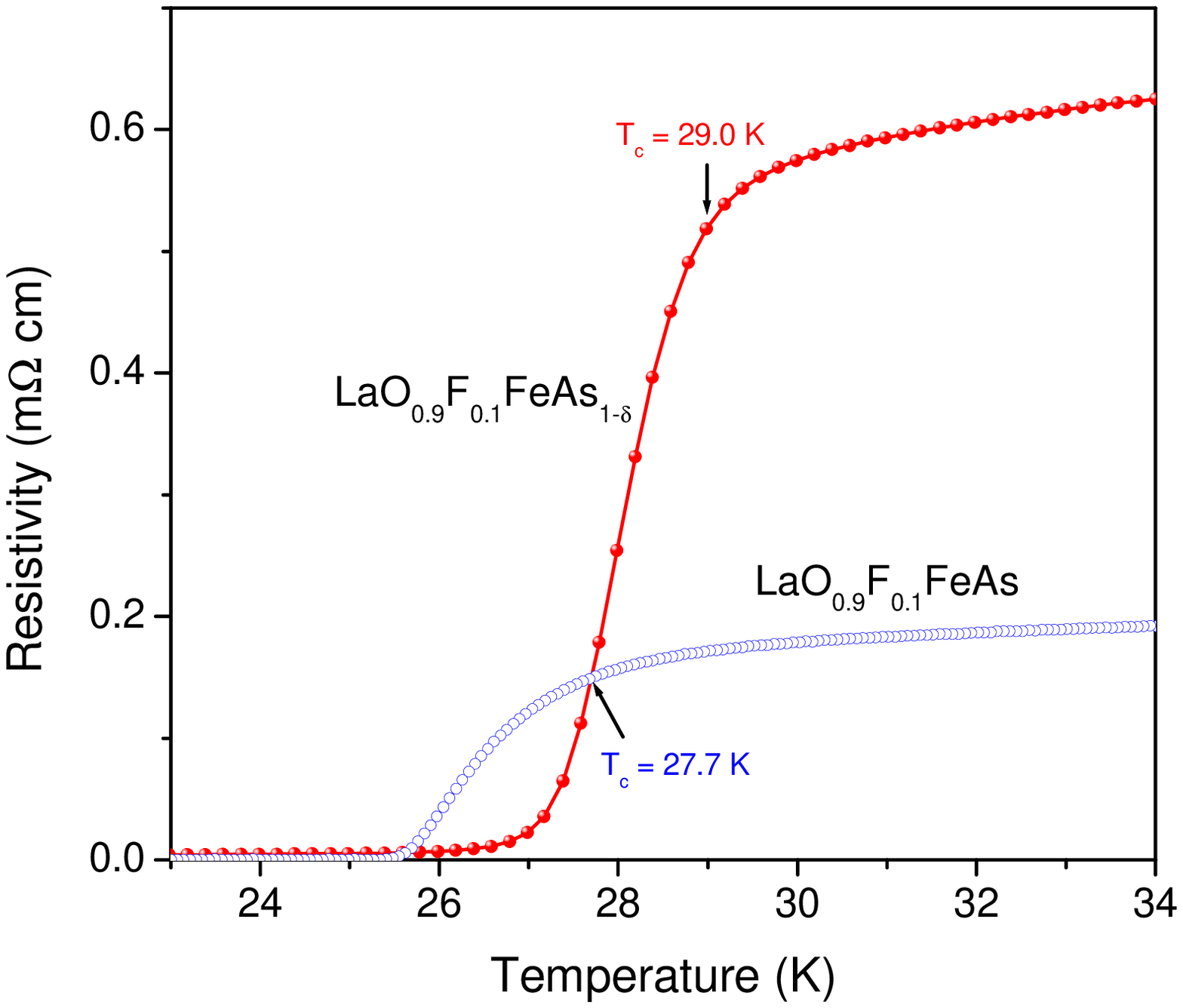}
\caption{Temperature dependence of the resistivity for the 
As-deficient (\textcolor{red}{$\bullet $}) and for the 
clean reference sample
(\textcolor{blue}{$\bullet $}) for temperatures up to 300~K (left)
and in the vicinity of $T_c$ (right).} 
\label{f2new}
\end{figure}
\begin{figure}[b]
\hspace{3cm} \includegraphics[width=10cm]{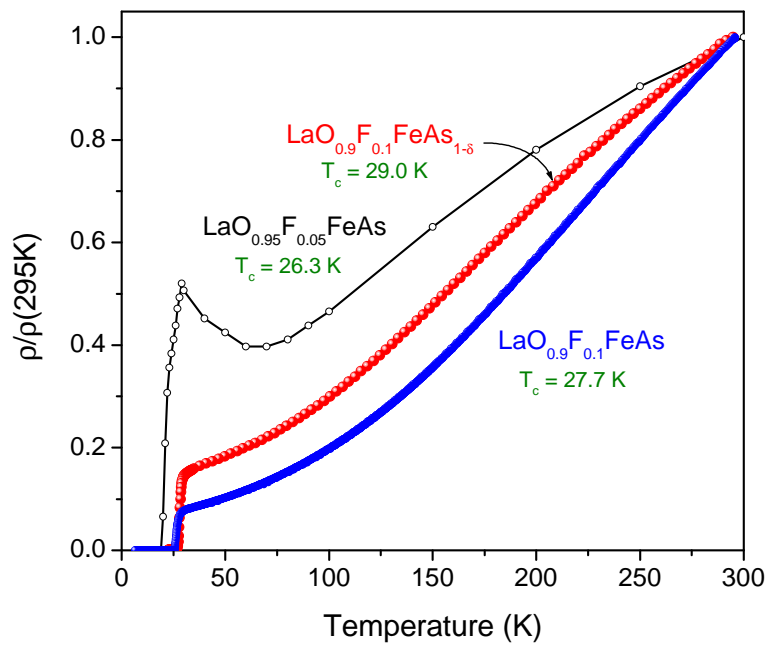}
\caption{$T$-dependence of the normalized resistivity for the 
As-deficient (\textcolor{red}{$\bullet$}),  the reference 
sample (\textcolor{blue}{$\bullet$}), and an underdoped sample ($\circ $)
taken from reference \cite{kohama-hall}.} 
\label{f5}
\end{figure}
 of the As-deficient sample 
in the normal state at 31~K, with about 0.6~$\mu \Omega$cm, exceeds 
that of the clean reference 
sample by a factor of about three. Since each As site is 
surrounded by four Fe sites, the effect of even a few As-vacancies might 
be drastic. Thus, a substantial shortening of the mean free path due to an 
As-deficieny of  about 0.1 seems to be quite reasonable. 
In spite of the resulting disorder in the FeAs layer, the As-deficient 
sample is found to exhibit, with $T_c = 29.0$~K, a {\it higher} 
transition
 temperature than the optimally doped reference sample ($T_c = 27.7$~K) 
and a relatively sharp transition width (see figure 2 (right)) which 
excludes an 
anomalous inhomogeneity. Compared with underdoped LaF$_{0.05}$O$_{0.95}$FeAs 
samples for which $T_c$-values of  26.3~K \cite{kohama-hall} 
and 20.6~K \cite{hess} 
were reported, the increase of $T_c$  
due to As-deficiency is even more pronounced. The unexpected 
increase of $T_c$ in the As-deficient sample might be caused by the 
suppression of the nesting related AFM
fluctuations  due to disorder effects and a possible additional 
non-phononic attractive
coupling induced by the localized vacancy related electronic states 
\cite{drechsler09} and/or by  the suppression
of pair-breaking interband scattering due to enhanced intraband scattering
within an As-vacancy stabilized $s_{\pm}$-scenario \cite{senga2}. 
The $\rho(T)$-dependence of the As-deficient sample 
resembles that of 
underdoped stoichiometric samples \cite{hess} 
only at high $T$ whereas it becomes similar to that of optimally doped 
samples at 
$T >T_c$  as shown in figure~3. 
In particular, the
pronounced low-temperature 
($T < 60$~K) upturn of $\rho(T)$ characterizing 
underdoped stoichiometric samples is not observed for 
our As-deficient samples. For more details see the data in 
supplementary part.

\subsection{Enhanced paramagnetism: Muon spin rotation measurements}
\begin{figure}[t]
\center{\includegraphics[width=0.7\columnwidth,angle=0,clip]
{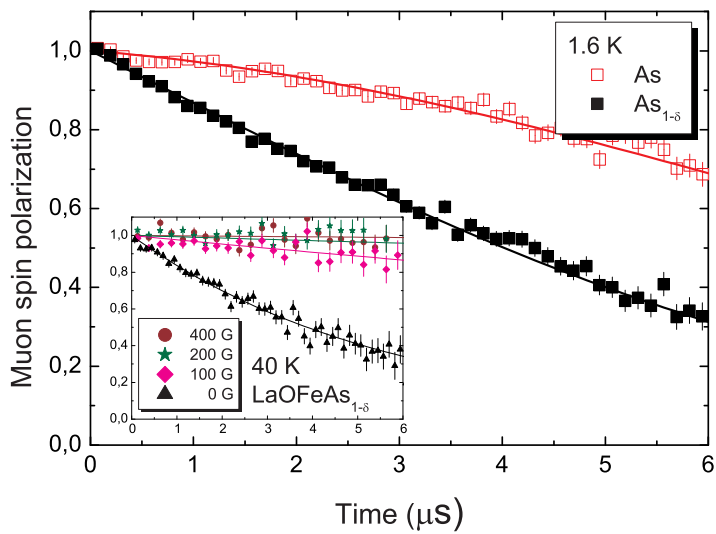}}
\caption[]{Zero field $\mu$SR spectra of LaO$_{0.9}$F$_{0.1}$FeAs
and LaO$_{0.9}$F$_{0.1}$FeAs$_{1-\delta}$ at 1.6~K. The inset shows
a longitudinal field experiment on
LaO$_{0.9}$F$_{0.1}$FeAs$_{1-\delta}$ at 40~K proving the static
nature of the weak electronic relaxation in this sample.}
\label{ZF-spectra}
\end{figure}
\begin{figure}[b]
\center{\includegraphics[width=0.7\columnwidth,angle=0,clip]
{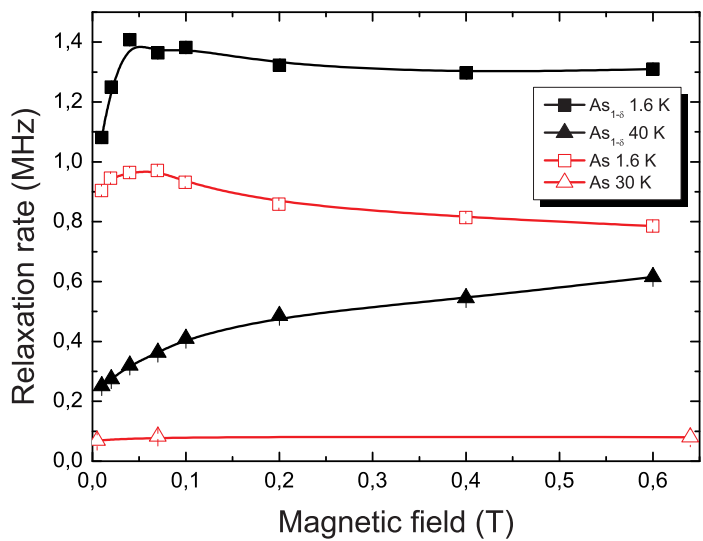}}
\caption[]{Field dependence of the transverse field $\mu$SR
relaxation rate of LaO$_{0.9}$F$_{0.1}$FeAs and
LaO$_{0.9}$F$_{0.1}$FeAs$_{1-\delta}$ in the normal and
superconducting state.} 
\label{TF-rates}
\end{figure}
The observed PLB of the upper
critical field at low $T$ and high external fields reported below
in section~4.2 should be caused by enhanced
paramagnetism. To confirm this presumption, 
we performed zero field (ZF) and transverse field (TF)
muon spin relaxation measurements ($\mu$SR) on our clean reference
and the As-deficient 
samples. In figure~\ref{ZF-spectra} we show ZF-$\mu$SR data at 1.6~K.
For the clean reference sample a weak Gaussian 
Kubo-Toyabe-like \cite{Hayano79} (KT)
decay  of the muon spin polarization is observed.
This relaxation can be traced back to the tiny magnetic fields
originating from nuclear moments. 
In contrast, for the As-deficient sample an
additional exponential relaxation due to electronic magnetic moments
is superimposed on the weak nuclear relaxation. Longitudinal field
(LF) experiments in the normal state at 40 K clearly prove a static
nature of the electronic relaxation (see inset in
figure~\ref{ZF-spectra}). Therefore we conclude that the disorder in
the As-deficient sample gives rise to the formation of dilute
quasistatic paramagnetic spin clusters. In a high external magnetic
field these spin clusters can give rise to additional internal
fields which reduce the upper critical field $B_{c2}$ as we have found 
experimentally.
An enhanced presence of paramagnetic electronic moments in the
As-deficient sample  is also clearly visible in the field dependence
of the TF-$\mu$SR relaxation rate depicted in figure~\ref{TF-rates}.
As well in the paramagnetic state (at 30 and 40~K, respectively) as
in the superconducting state at 1.6~K the relaxation rate for the
As-deficient sample is much stronger than for the nominal
composition. In particular, the increase of the relaxations rate
with increasing field is typical for a paramagnetic system.

\section{Upper critical field}
\subsection{Resistance for applied static and pulsed fields} 

In figure~6 the electrical resistance of the studied As-deficient sample 
\begin{figure}[b]
\hspace{2.5cm}
\includegraphics[width=10cm]{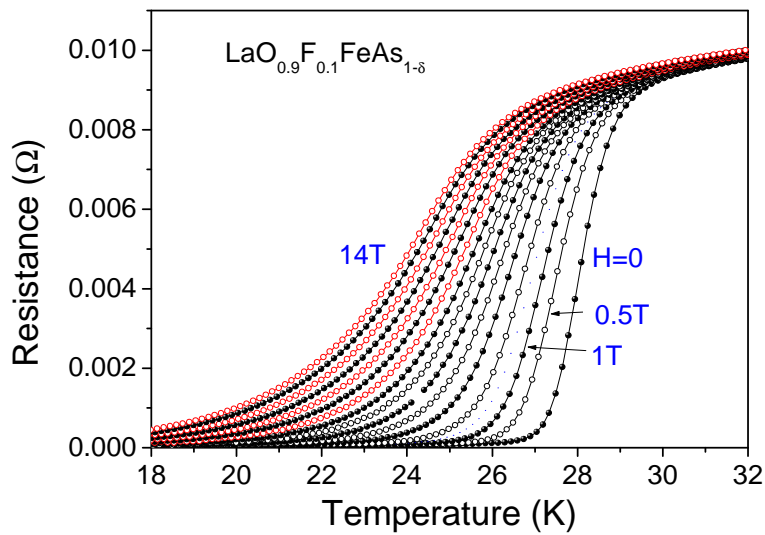}
\caption{The $T$-dependence of the resistance $R$ for the 
As-deficient sample
for various dc fields up to 14~T. Between 1 and 14~T, the applied magnetic 
field was increased in steps of 1~T.} \label{f9}
\end{figure}
\begin{figure}[t]
\hspace{1.5cm}\includegraphics[width=12cm]{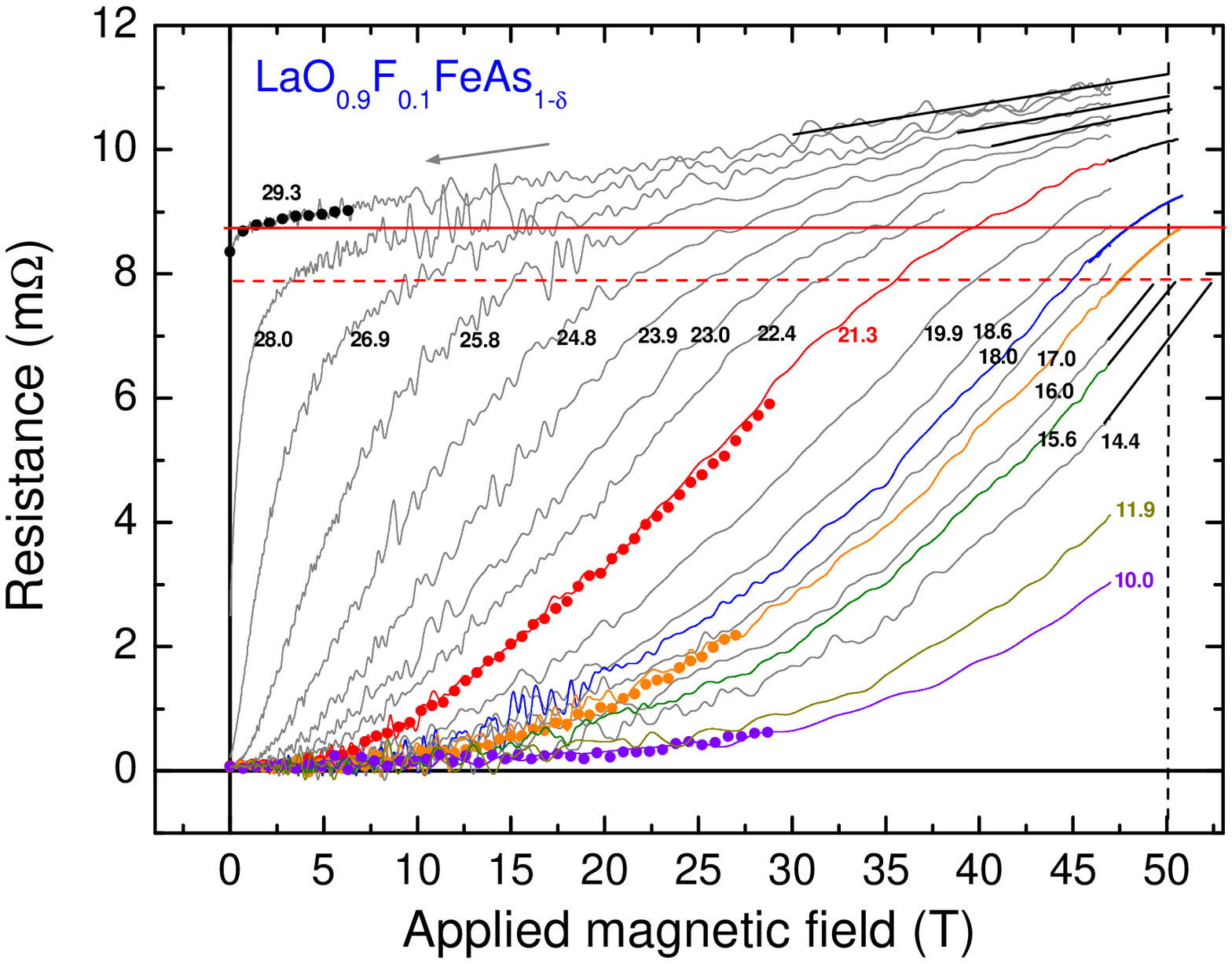}
\caption{Field dependence of the resistance at 
fixed $T$ (see legend) measured in pulsed fields.  Lines: 
measurements up to 47~T; symbols: measurements up to 29~T 
shown for four selected 
$T$-values. Horizontal full and dashed lines: $R=R_N$ and
$R=0.9R_N$, respectively, where $R_N$ is the field-independent resistance
in the normal state obtained by an extrapolation of $R(H)$ at 29.3~K to $H=0$.} \label{f10}
\end{figure}
is plotted against the temperature for applied dc fields up to 
14~T. With 
increasing applied field, the onset of superconductivity is found to 
be
shifted to lower temperatures. Additionally, a substantial 
broadening of the transition curves is observed at high applied fields 
which mainly stems from the large anisotropy of the upper critical field 
$B_{c2}$ which is expected for the layered Fe oxypnictide superconductors
 \cite{mazin1,singh}.

We performed measurements in pulsed high magnetic fields up to 60~T in 
some cases.  
Resistance data obtained in fields up to 50~T for the As-deficient sample 
are plotted in figure~7. Gold 
contacts (100~nm thick) were made by sputtering in order to provide a 
low contact resistivity and, therefore, to avoid possible heating effects 
in the 
high-field measurements 
which might seriously distort the shape of the transition curves
affecting the $B_{c2}$-values to be derived. 
The magnetic field generated by the employed IFW's pulsed field 
magnet rises within about 10~ms to 
its maximum value $B_{max}$ (which can be varied up to 50~T) and 
decreases afterwards
to 
zero within the same time. The resistance data shown in figure~7 were taken 
for 
descending field using $B_{max} = 47$~T. Additionally, resistance data were 
collected for $B_{max} = 29$~T at several selected temperature. The agreement 
between the resistance data for $B_{max} = 47$~T and 29~T confirms that 
our data are not affected by sample heating. 
	Again, a pronounced broadening of the transition curves is observed 
at high magnetic fields which is associated with the large anisotropy 
$ \gamma_H = B_{c2}^{ab}/B_{c2}^c$ . 
For polycrystalline samples, only the higher 
$B_{c2}^{ab}$ is accessible. Since it
is related to those grains 
oriented with their $ab$-planes along the applied field,
$B_{c2}^{ab}$ 
 can be determined 
from the onset of superconductivity. The onset of first dissipation in the 
resistive transition curves can be roughly associated with grains of the 
lower $B_{c2}^c$. However, one has to taken into account that the estimation 
of the anisotropy $\gamma_H = B_{c2}^{ab}/B_{c2}^c$ from resistance 
measurements provides only a lower limit of $\gamma_H$. 
Adopting a simple Ginzburg-Landau picture with mass anisotropy,
values of $\gamma$  predicted from the local-density approximation (LDA)
were found to vary between 6.2 and 15 \cite{mazin1,singh,drechsler09}
whereas those from magnetic torque measurements on SmFeAsO$_{0.8}$F$_{0.2}$ 
and NdFeAsO$_{0.8}$F$_{0.2}$ single crystals  
have been found between about 7 (at $T_c$) and 19 
(for $T \rightarrow 0)$ \cite{weyeneth}.

\subsection{Different criteria for the determination of $B_{c2}$}	

The $R(H)$-curves in figure~7 reveal a considerable magneto-resistance 
of the investigated sample at high magnetic fields.  
In a first approach, the 
upper critical field $B_{c2}^{ab}$ was determined as in reference 
\cite{hunte} from the onset of superconductivity defining it at 90~\% of the 
resistance $R_N$ in 
the normal state and ignoring the magneto-resistance. 
This criterion corresponds to the dashed horizontal line in figure~7. 
The $B_{c2}^{ab}(T)$-curve of our As-deficient sample obtained for the 
so defined upper critical field is shown in figure 8 together  with the
data found for our clean
reference sample and for another clean sample
reported by Hunte {\it et al.} \cite{hunte}.
\begin{figure}[t]
\hspace{2.5cm}
\includegraphics[width=10cm]{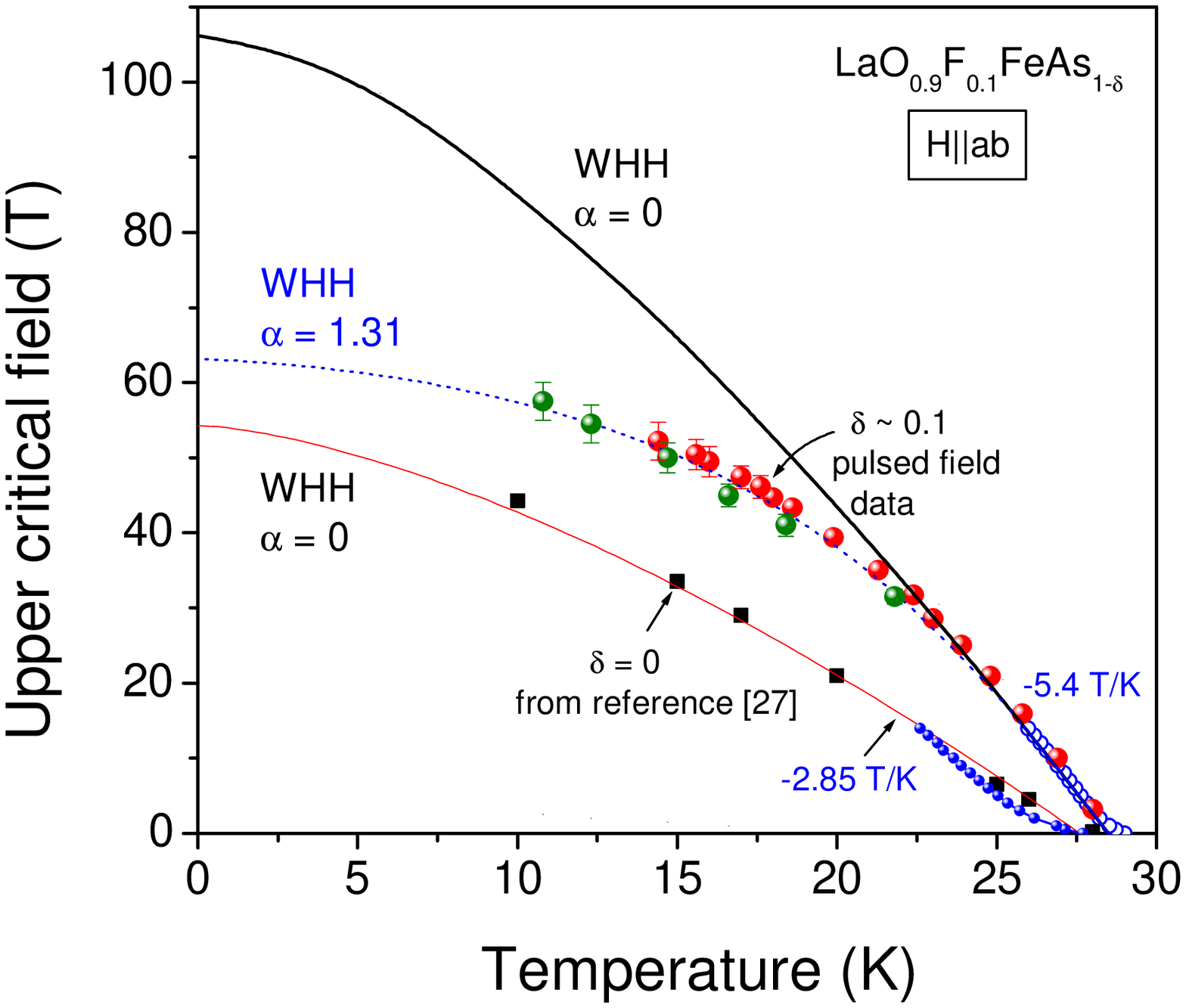}
\caption{$B^{ab}_{c2}$ vs~$T$. Data for the As-deficient sample
from DC ($\textcolor{blue}{\circ} $
) and pulsed field measurements (
$\textcolor{red}{\bullet} $- IFW Dresden,
$\textcolor{green}{\bullet} $- FZD). 
$\textcolor{blue}{\bullet} $ and $\blacksquare$
- data for our and another clean reference samples, respectively.
The latter are taken from reference \cite{hunte}. 
Solid lines: WHH-model 
without Pauli-limiting. Dotted line: $B_{c2}(T)$ for $\alpha$~=~1.31 
without spin-orbit scattering, 
where $\alpha$ is the Maki parameter - see equation (6).}
\label{f11}
\end{figure}
\begin{figure}[b]
\hspace{0.1cm}
\begin{minipage}{7.9cm}
\includegraphics[width=7.85cm]{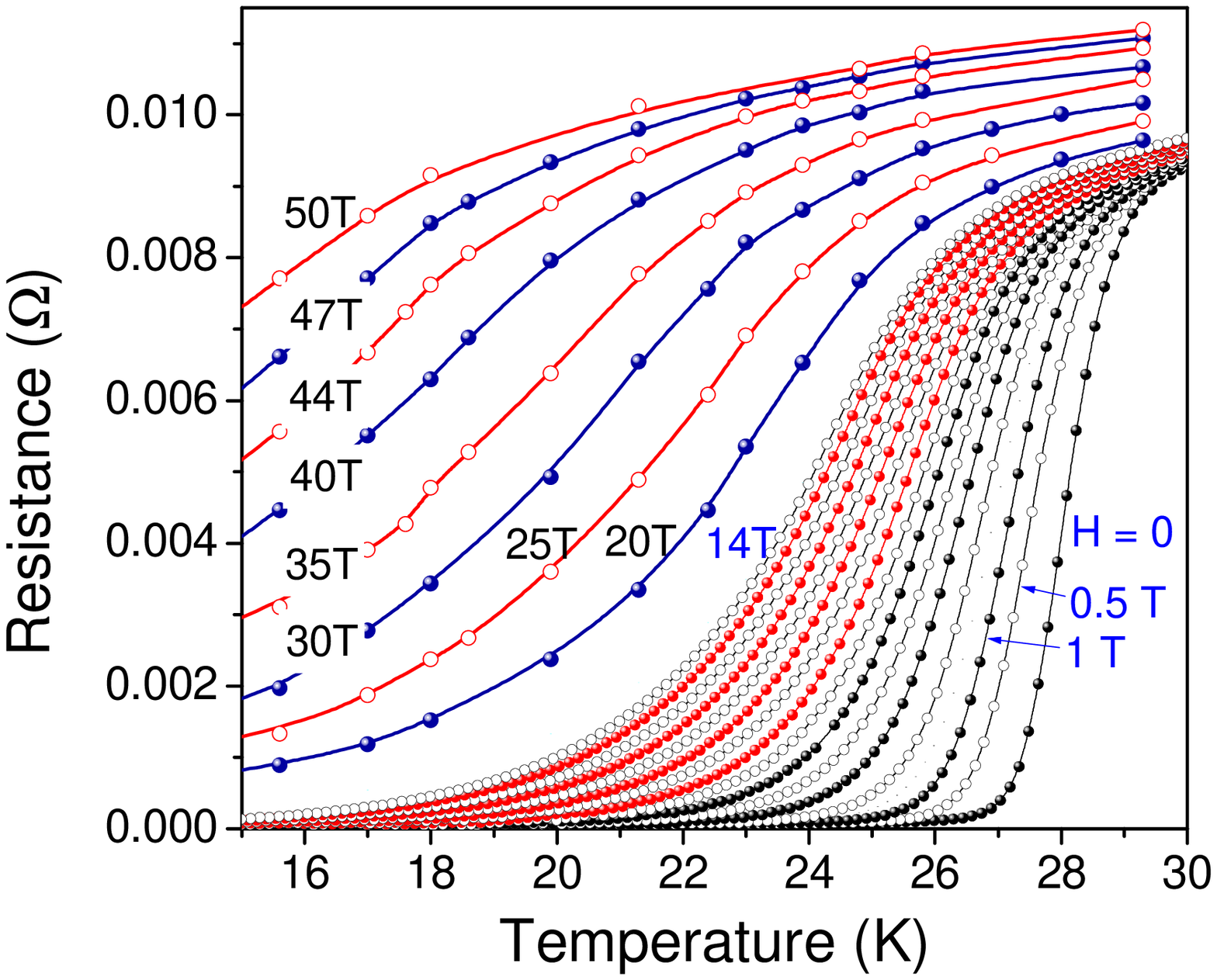}
\end{minipage}
\hspace{0.3cm}
\begin{minipage}{7.8cm}
\includegraphics[width=7.2cm]{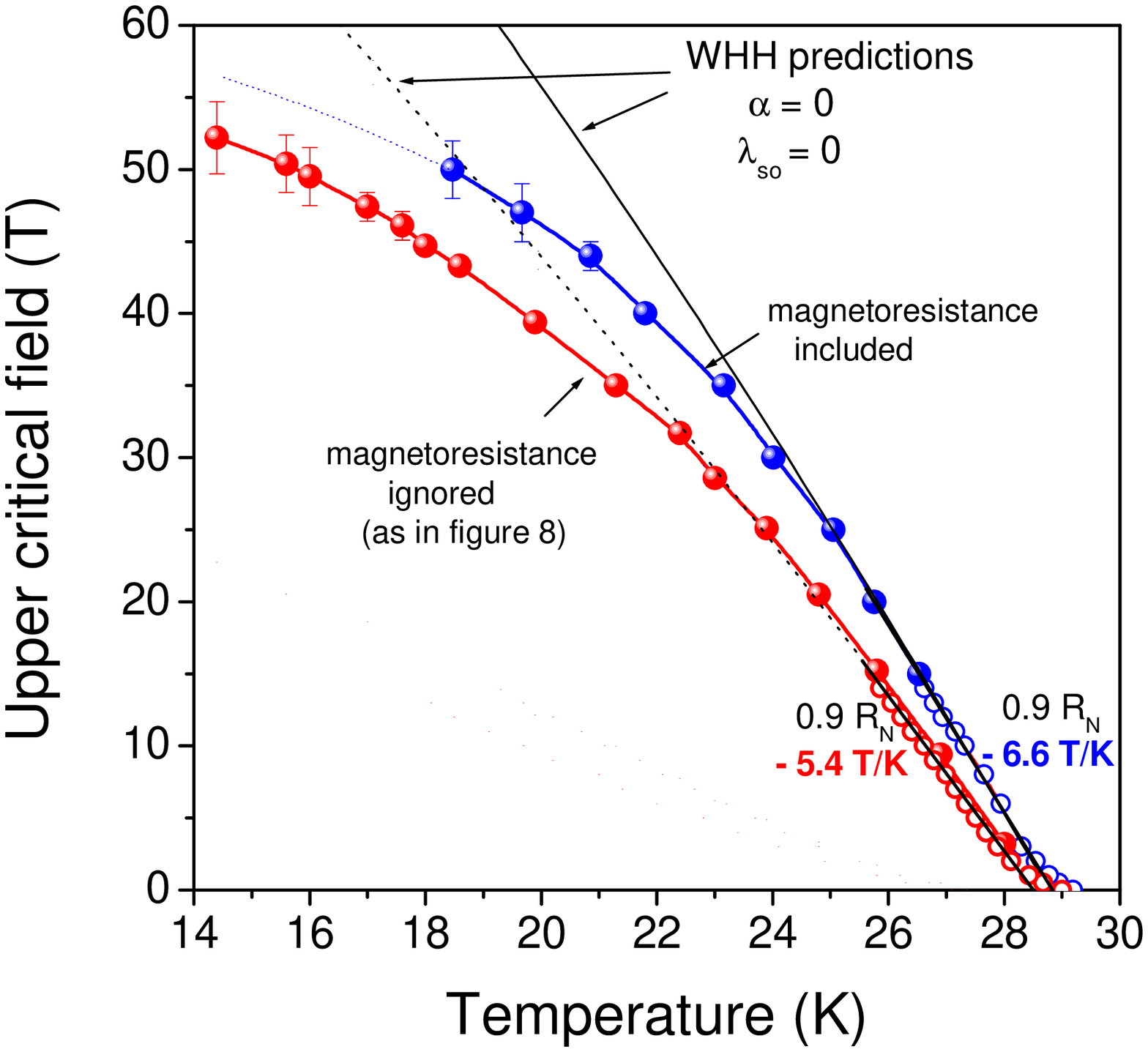}
\end{minipage}
\caption{The $T$-dependence of the resistance from both dc and pulsed 
field data
(left). Upper critical field versus $T$ from 90\%  
of the normal state resistance $R_N$ from different 
$B_{c2}$-definitions (right). 
Open (filled) symbols: dc 
field data - figure~6 (pulsed field data - figure~7).} 
\label{f12}
\end{figure}
For our As-deficient sample, good agreement between dc and pulsed field 
measurements is obtained in the 
field range up to 14~T. The corresponding $B_{c2}(T)$-curve in figure 8 
shows a surprisingly steep $dB_{c2}/dT \mid _{T_c} = - 5.4$~T/K which 
exceeds the slopes of $B_{c2}(T)$ of the two clean samples 
by more than a factor of two. This points to strong impurity scattering in 
the As-deficient sample in accord with its enhanced resistivity at 30~K. 
For the clean sample \cite{hunte} the available $B_{c2}$-data up to 45 T
is well described by the WHH-model \cite{WHH} assuming that the $B_{c2}(T)$
is limited by orbital effects, only (see figure 10 (left)). Whereas for the 
As-deficient sample the WWH-model fits the experimental data up to 
about 30 T, only.
Using  $dB_{c2}/dT = - 5.4$~T/K and $T_c = 28.5$~K, 
this model predicts $B^*_{c2}(0) = 0.69 T_c (dB_{c2}/dT)_{T_c} = 106$~T at 
$T =$0. However, for applied fields above 30~T or at temperatures below 23~K, 
increasing deviations from the WHH-curve are clearly visible both for the 
$B_{c2}$-data from the IFW~Dresden and the FZD. The small difference
between both data sets is within the error bars of both measurements.
The flattening of $B_{c2}(T)$ 
at high fields points to its limitation by the Pauli spin paramagnetism 
as will be discussed in more detail in the next section. 
	In order to check, whether the observed deviations of the 
experimental $B_{c2}(T)$ data from the WHH prediction are affected by the 
definition of the upper critical field, within a second approach in 
defining $B_{c2}$, the magnetoresistance in the normal state and the 
temperature dependence of $R_N$ were taken into account. The resistance 
data vs.\ temperature plotted in figure~9 (left) for the 
As-deficient sample are taken
both from dc and pulsed field measurements. Within 
this approach, $B_{c2}^{ab}$ was defined at 90\% of the resistance $R_N(T,H)$ 
in the normal state. The temperature dependence of $R_N$ was approximated 
by $R_N(T) = 7.74 + 1.7\cdot 10^{-2}T^{1.4}$ where $R_N$ 
is given in m$\Omega$ and $T$ in K. This relation was found to fit the 
experimental $R_N(T)$ data between $T_c$ and 80~K very well. For this 
modified definition of $B_{c2}$ , one gets somewhat higher $B_{c2}$-values 
than for the first one as shown in figure~9 (right). The slope 
$dB_{c2}/dT\mid_{T_c}$
 becomes, with  $dB_{c2}/dT \mid_{T_c} = - 6.6$~T/K even steeper 
resulting in an enhanced field  $B^*_{c2}(0) = 131$~T at $T = 0$ 
predicted by the WHH model. More importantly,  the resulting 
difference between the measured $B_{c2}(T)$ and 
the extrapolated $B^*_{c2}(T)$ at lower $T$ is comparable 
for both definitions of the upper critical field. Hence, the flattening of 
the experimental $B_{c2}(T)$-curve observed at high magnetic fields is rather 
similar, regardless of whether the magnetoresistance is taken into account 
in defining $B_{c2}$ or not.

\section{\bf Analysis of $B_{c2}$ and discussion}

\subsection{Orbital and paramagnetic upper critical field}

Standard superconductivity 
as described by the BCS- or the more sophisticated Eliashberg-theory
rests on Cooper-pairs.
They consist of two electrons with 
opposite spins (non-$p$-wave) and momenta. 
Hence, there are two 
magnetic channels to affect a superconducting state: (i) 
by the Lorentz-force acting via the charge and the opposite momenta 
(the phases) on the paired 
electrons (usually called diamagnetic or orbital effect) and (ii) the spin 
channel (called also paramagnetic effect) where a
singlet pair is transfered in a practically unbound triplet,
i.e.\ it is broken by the Zeeman effect. 
Orbital and spin pair breaking (of a singlet Cooper pair) in the presence of a 
magnetic field are illustrated in figure~10 (left).
\begin{figure}[t]
\includegraphics[width=7.5cm]{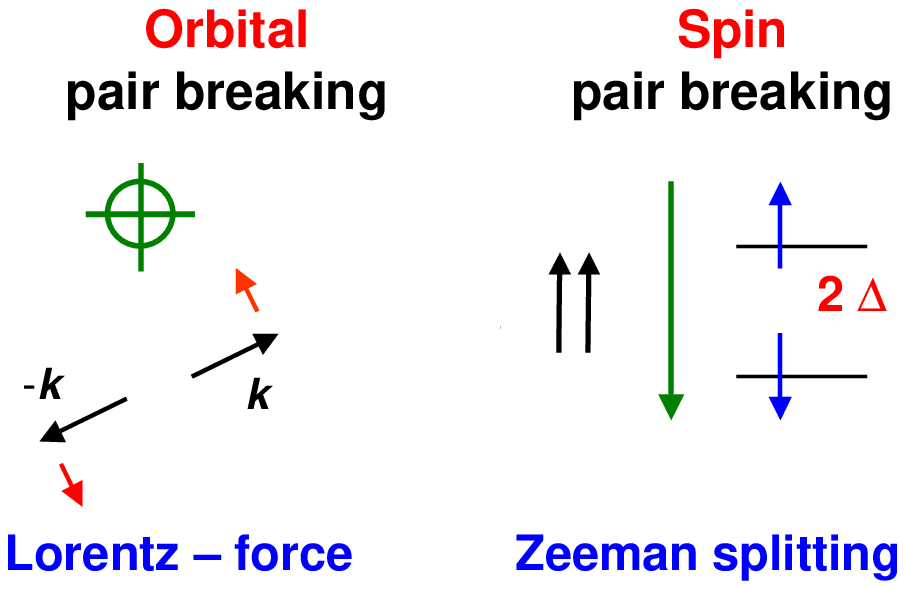}
\hspace{2cm}
\includegraphics[width=5.5cm]{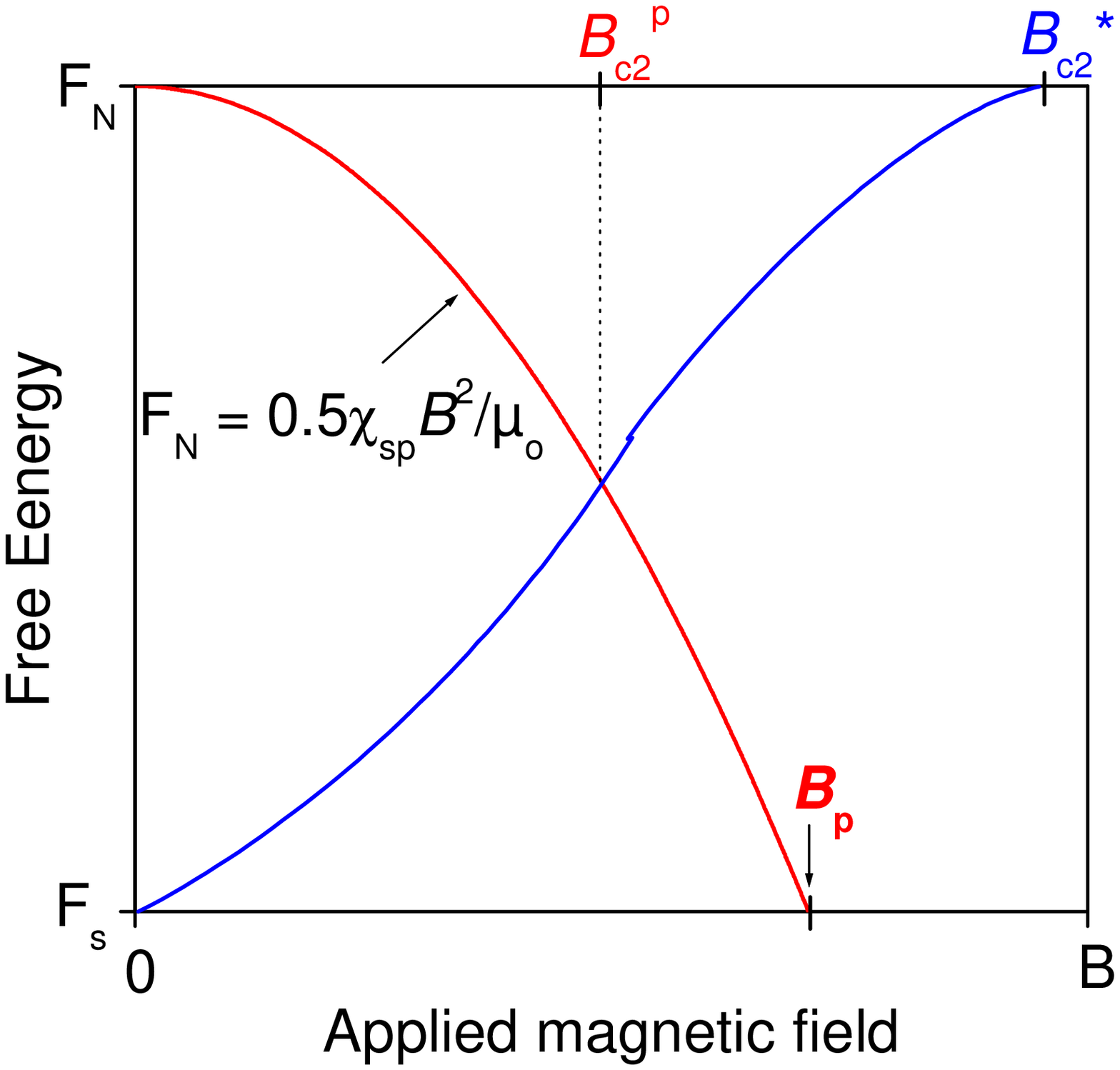}
\caption{
Schematical view of the magnetic field affecting a Cooper-pair (left). 
Pair breaking due to the Lorentz force acting via the charge on the 
momenta of the paired electrons ({\it orbital} pair-breaking) or due to 
the Zeeman effect aligning the spins of the two electrons with the applied 
field ({\it spin} pair-breaking). Field dependence of the Gibbs-free energy 
schematically (right). The free energy in the normal state
with Pauli spin susceptibility $\chi_{sp}$ (parabolic 
red line) crossing the zero-field free energy in the superconducting state 
at the Pauli limiting field $B_p$ and without $\chi_{sp}$  
(upper horizontal line). Blue curve: Free energy of a type-II 
superconductor crossing the normal state at $B_{c2}^*$ or $B_{c2}^p$. 
} 
\label{f13}
\end{figure}
Usually,
at high temperature below $T_c$, the suppression in the orbital channel is more
effective. Then the  paramagnetic effects may be visible at low-temperature and
high fields, only. Thus, at 
sufficiently large magnetic field,
the superconductivity is destroyed  
by orbital {\it and} spin pair breaking.
(Contrarily, in the hypothetical case of bipolarons $B_{c2}(0)$ is essentially
unlimited or extremely large \cite{alexandrov}.)

According to the WHH approach 
\begin{equation}
B^*_{c2}(0) = 0.69 T_c (dB_{c2}/dT)_{T_c} \quad ,
\end{equation}
the orbital limited upper critical field $B^*_{c2}(0)$
 is related both to $T_c$ and to the slope of $B_{c2}(T)$ at $T_c$. 
The relation between the orbital and spin paramagnetically limited 
upper critical fields is illustrated in figure~10 (right), where the 
Gibbs 
free energy 
is plotted against the applied magnetic field. 
By applying a magnetic field, 
the free energy of the superconductor in the superconducting state, $F_s$, 
increases, whereas its free energy in the normal state, $F_N$, is lowered by an 
amount of 0.5$\chi_{sp} B^2/\mu_0$ with $\chi_{sp}$ being
 the Pauli spin susceptibility. The field at 
which $F_N(B)$ becomes equal to the condensation energy of the superconductor 
defines the Pauli limiting field $B_p$ which is for weakly coupled 
superconductors given by 
\begin{equation}
\label{Bp}
B_p(0) [\mbox{Tesla}] = 1.86 T_c\mbox{ [K]} \sqrt{1+\lambda } = 
1.06\Delta_0  \mbox[K] \sqrt{1+\lambda } \ ,
\end{equation}
at $T = 0$, as was pointed out by Clogston \cite{clogston} 
 and Chandrasekhar \cite{chandrasekhar}, where $2\Delta_0$ denotes 
(the isotropic) gap in the adopted there $s$-wave picture. 
Including strong coupling corrections to 
BCS  due to {\it el-boson} and {\it el-el} 
interactions, one gets 
\begin{equation}
B_p(0) \ \mbox{[Tesla]} = 1.86 \eta_{\Delta}\eta_{eff}(\lambda)  T_c  \mbox{[K]}
=1.06\Delta_0  \mbox[K] \eta_{eff}(\lambda) \ , 
\label{}
\end{equation}
where $\eta_{\Delta}$ describes the strong
coupling intraband correction for the gap, 
$\eta_{eff}(\lambda) = (1+\lambda)^\varepsilon  
\eta_{ib} (1-I )$ with $I$ as the Stoner factor $I=N(0)J$
\cite{orlando,schossmann}, $N(0)$ is the electronic density
 of states (DOS) per spin at the Fermi level $E_F$, $J$ is an
effective exchange integral, and $\eta_{ib}$ has been introduced to describe
phenomenologically the effect of  
gap anisotropy, multi-band character, energy dependence of states 
etc., which are possibly present
also in FeAs based superconductors (see
for instance the references \cite{hans,dubroka,dias1}). 
It is assumed to be strong enough to compensate the effect of the 
unknown Stoner factor.
The so called strong coupling (i.e.\ finite $el$-$boson$ coupling)
correction factor $\eta_{eff}(\lambda)$ scales with 
$(1+\lambda)^{\varepsilon}$, where the exponent $\varepsilon$ amounts either
0.5 or 1 according to references 
\cite{orlando,schossmann}, respectively.
The paramagnetically 
limited upper critical field, $B^p_{c2}$,  corresponds to that field at 
which $F_N(B)$ and $F_s(B)$ are equal. $B^p_{c2}$ is always lower than both 
$B^*_{c2}$ and $B_p$ as illustrated in figure~10 (right). 
According to Maki \cite{maki}, the paramagnetically limited field 
$B^p_{c2}$ reads
\begin{equation}
B^p_{c2} (0) = B^*_{c2}(0)/\sqrt{1 +\alpha^ 2}				
\end{equation} 
where the Maki parameter, $\alpha$,  is given by
\begin{equation}
\label{alpha}
\alpha = \sqrt{2} B^*_{c2}(0)/ B_p(0) \ .
\end{equation}
The Maki parameter $\alpha$ provides a convenient 
measure for the relative strength of 
orbital and spin pair-breaking. Within the WHH approach
the shape of $B_{c2}(T)$ depends sensitively on the magnitude of $\alpha$,
namely, with increase of $\alpha$ an increasing  
flattening of the $B_{c2}(T)$-curve is predicted. Introducing a second
auxilary parameter $\lambda_{s0}$ ascribed to spin-orbit scattering 
the strong effect of $\alpha$ could be partially
reduced. Anyhow, since in our As-deficient sample the effect
of spin-orbit scattering on $B_{c2}(T)$ is expected to be rather 
weak, only,
it has been ignored in our analysis \cite{remark-spin-orbit}.  

\subsection{As-deficient LaO$_{0.9}$F$_{0.1}$FeAs$_{1-\delta}$}

In figure~8, the $B_{c2}(T)$ data of our As-deficient 
samples are analyzed within the WHH model.
A satisfying fit of the experimental
 data to this model was obtained for $\alpha = $1.31. Using 
$B_{c2}^*(0) = $106~T (and 131~T in the second approach to determine 
$B_{c2}$ in which the magneto-resistance was taken into account) one 
obtains $B_{c2}^p (0) = 63$~T (and 68~T within the second approach) for the 
upper critical field at $T = 0$ and $B_p(0) = $114~T (and 141~T 
within the second
 approach) for the Pauli limiting field from equations (5) and  (6), 
respectively. We used $\lambda > $0.6 \cite{drechsler,drechsler09}
(estimated for the clean sample with a bulk $T_c$ of about 26~K
\cite{luetkens})
for a 
representative value of the {\it el-boson} coupling constant for
 LaO$_{0.9}$F$_{0.1}$FeAs$_{1-\delta}$ and $T_c = $28.5 K 
and estimated a rather large value of $\eta_{\Delta}\eta_{eff}  = $2.09 
from equation (4). The dotted $B_{c2}^p (T)$ line plotted in figure~8 
is based on equation (5) and was obtained by replacing $B_{c2}^*(0)$ 
entering both its numerator and $\alpha$ by $B_{c2}^*(T)$ of the 
WHH model for $\alpha = 0$. This rough approximation of $B_{c2}^p (T)$ has 
been used to illustrate the $T$-dependence of the upper critical 
field due to PLB in the studied As-deficient samples.

\subsection{Comparison with other samples: slope of $B_{c2}(T)$ near 
$T_c$, disorder, and paramagnetism} 

Superconductivity in $R$OFeAs  ($R$= La, Pr, Sm, Nd, Gd) can be induced by 
carrier doping due to the suppression of the magnetic order and the 
structural phase transition observed for the parent compounds. This can be 
done by substituting F for O (in the case of $R$= La, Pr, Sm, Nd, Gd), by 
partial removing of O (for $R$=  Sm, ) or, in the case of GdOFeAs and 
LaOFeAs, by substituting Th for Gd and Sr for La, respectively. High 
values of 
$T_c = 55$~K have been reported both for 
SmO$_{0.9}$F$_{0.1}$FeAs and 
SmO$_{0.85}$FeAs. A slightly higher $T_c$
 of 56 K was reported for Gd$_{1-x}$Th$_{x}$OFeAs. 
Superconductivity can be obtained also by direct carrier doping 
into the conducting FeAs planes which are essentially for superconductivity 
substituting Co 
\cite{yamamoto, sefat-Co-doping,sefat-Co-doping-BaFe2As2,leithe-jasper-Co-doping,
Matsuishi-Co-doping-CaFFeAs,matsuishi-Co-doping-SrFe2As2,Cao-Co-doping} 
(or Ni \cite{Cao-Ni-doping,Li-Ni-doping-BaF2As2,ren-Ni-doping-EuFe2As2} 
or Zn \cite{Li-Zn-doping})
 for Fe, as was demonstrated for 
LaOFeAs, CaFFeAs,
 BaFe$_2$As$_2$, and EuFe$_2$As$_2$ . 
Replacing Fe with Co (or Ni) is expected not only to 
carrier doping, but also to introducing disorder in the FeAs layer. It is 
remarkable that these superconducting compounds can tolerate considerable 
disorder in the FeAs layers. 

In compounds with enhanced disorder in the FeAs layers, a strong increase of 
$B_{c2}(T)$ and its slope near $T_c$ was found.
We suggest that this results from a reduced mean free path and an
 enhanced intraband scattering like in usual dirty $s$-wave superconductors
such as NbTi \cite{larbalestier}. For instance, a relatively large slope of 
d$B_{c2}/$d$T = - 4.9$~T/K near $T_c$ has been  reported for 
the Co-doped Ba(Fe$_{0.9}$Co$_{0.1}$)$_2$As$_2$ system \cite{yamamoto}, 
which is only slightly below the value of -5.4 T/K  for our As-deficient 
La-1111 sample. Disorder due to As vacancies seems to be responsible 
also 
for the large slope of d$B_{c2}/$d$T = - 6.3$~T/K near $T_c$ reported for 
(Ba$_{0.55}$K$_{0.45}$)Fe$_2$As$_2$ \cite{altarawneh} as will be discussed 
below in more detail. The above mentioned slopes of 
$B_{c2}(T)$ near $T_c$ for these somehow disorded systems are  
included in table~1. For completeness we note that the rare earth-1111 
systems exhibit large slopes d$B_{c2}/$d$T \stackrel{<}{\sim} -9$~T/K 
\cite{jaroszynski,jia}. In some cases also an onset of a flattening like 
for our sample has been observed \cite{jaroszynski} (see figure 11 right and 
the discussion below). In our opinion this should be caused 
due to paramagnetic effects of unkown microscopic origin.   
A remarkable 
relatively large isotropic slope of -5.96~T/K 
for both directions, $\parallel$ to  the basal plane and $\perp$ it 
at a low $T_c$ of 
12.4 K has been reported also for a Fe$_{1.03}$Te single crystal 
\cite{chen-FeTe} caused by {\it excess} Fe ions (i.e.\ excess Fe impurities 
at minor Fe(2) positions).
The weak anisotropy of about 1.6 is also noteworthy.   
\begin{figure}[b]
\includegraphics[width=7.5cm]{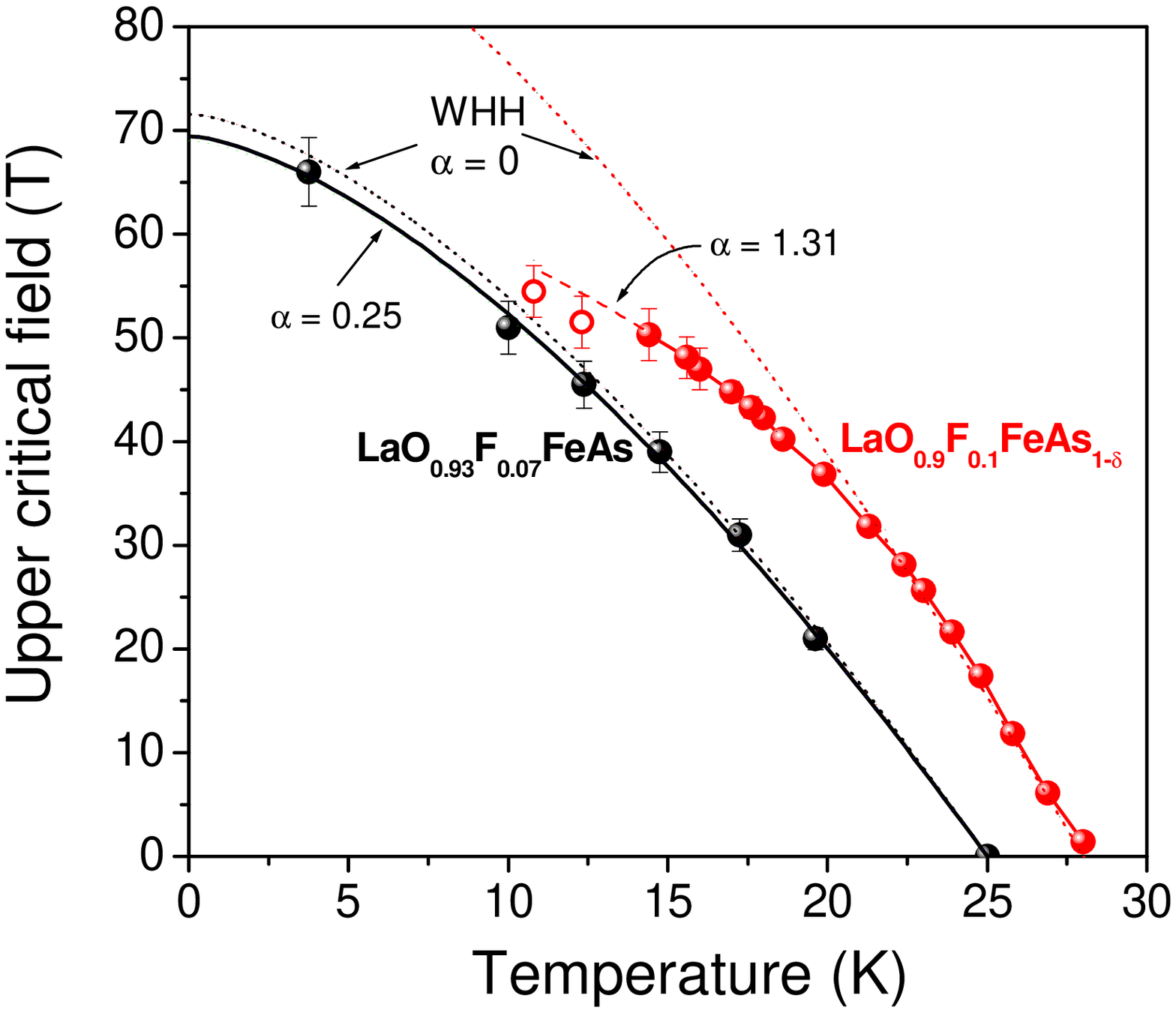}
\includegraphics[width=7.35cm]{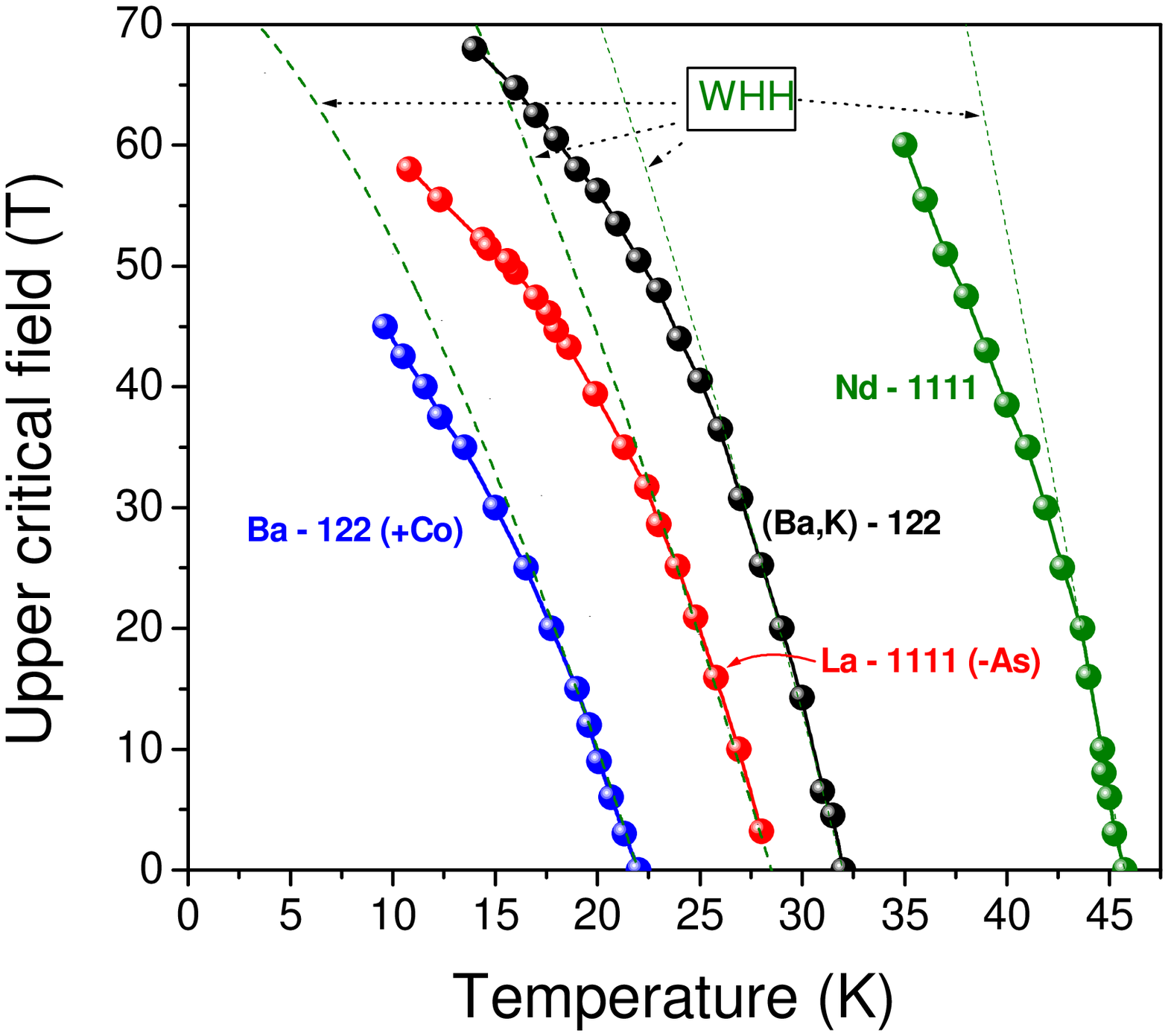}
\label{f14}
\caption{
$B_{c2}$ vs $T$-data of an As-deficient 
LaO$_{0.9}$F$_{0.1}$FeAs$_{1-\delta}$ sample compared with data 
reported by Kohama {\it et al.} \cite{kohama-comp}
for a non-deficient LaO$_{0.93}$F$_{0.07}$FeAs sample (left). In both cases,
 $B_{c2}$  is defined at 0.8 $R_N$ according to the criterion used in 
reference \cite{kohama-comp}. Comparison of $B_{c2}$ vs $T$-data of various 
disordered Fe-pnictide superconductors (right). In all cases, $B_{c2}$ is 
defined at 0.9$R_N$. \textcolor{red}{$\bullet$}: our As-deficient 
LaO$_{0.9}$F$_{0.1}$FeAs$_{1-\delta}$ sample. The data for 
Ba(Fe$_{0.9}$Co$_{0.1}$)$_2$As$_2$ (\textcolor{blue}{$\bullet$}), 
Ba$_{0.55}$K$_{0.45}$Fe$_2$As$_2$ ($\bullet$) 
and NdO$_{0.7}$F$_{0.3}$FeAs single crystals (\textcolor{green}{$\bullet$}) 
were taken from references \cite{yamamoto,altarawneh,jaroszynski}. 
Dashed lines: WHH model for $\alpha = 0$. All curves shown  correspond to 
the  $H \parallel (a,b)$ case.} 
\label{f14}
\end{figure}
Another puzzling observation to be understood is the fact that 
many of the known FeAs-based SCs show almost no 
Pauli-limiting behaviour up 
to 70~T as examined at present including even systems with relatively low 
$T_c$-values. For example, the $B_{c2}(T)$ data reported for 
LaO$_{0.93}$F$_{0.07}$FeAs \cite{kohama-comp}
only slightly deviate from the WHH curve 
as shown in figure~11~(left). In contrast, few more or less strongly disordered 
systems exhibit clear deviations from the WHH curves, qualitatively similar 
to our findings reported above, see figure~11~(right). 

For the As-stoichiometric 
reference family LaO$_{1-x}$F$_x$FeAs it has been reported that the Pauli 
paramagnetic susceptibility within  this series 
 is affected by the  
F doping showing a maximum around 
$x=0.05$ \cite{nomura}.
We found for our As-deficient samples also indications for
a strongly enhanced Pauli paramagnetism from 
$\mu$SR experiments as was discussed above. 
This explains the flattening of $B_{c2}(T)$ observed for this sample at 
applied fields above 30~T.    
	We analysed the $B_{c2}(T)$ data shown in figure~11 in order to 
determine the different strength of the
paramagnetic pair-breaking in these 
samples. For LaO$_{0.93}$F$_{0.07}$FeAs (see figure~11~(left)), a 
small value of the
Maki-parameter of $\alpha  = $0.25 is derived. 
A sizable paramagnetic pair-breaking effect is 
expected for larger values of the Maki-parameter, 
i.e.\ according to 
equation (6) for large orbital $B_{c2}^*(0)$ and/or low Pauli-limiting 
fields $B_p(0)$. Indeed, the  value of the As-deficient sample is, 
with $\alpha = $1.31,  by more than a factor of 5 larger than  estimated 
for the LaO$_{0.93}$F$_{0.07}$FeAs sample. This is mainly due to the low 
Pauli limiting field estimated for the As-deficient sample which is, 
with $B_p(0) = $114~T almost three times smaller than $B_p(0)$ estimated 
for LaO$_{0.93}$F$_{0.07}$FeAs 
(see table~1). The low value of $B_p(0)$ of the 
As-deficient sample can be explained by its enhanced Pauli spin 
susceptibility. Additionally, its large orbital $B_{c2}^*(0)$ contributes to 
the observed Pauli-limiting behaviour of this sample which is enhanced by 
the large values of both $T_c$ and $dB_{c2}/dT$ at $T_c$.
\begin{figure}[t]
\hspace{2.5cm}
\includegraphics[width=10cm]{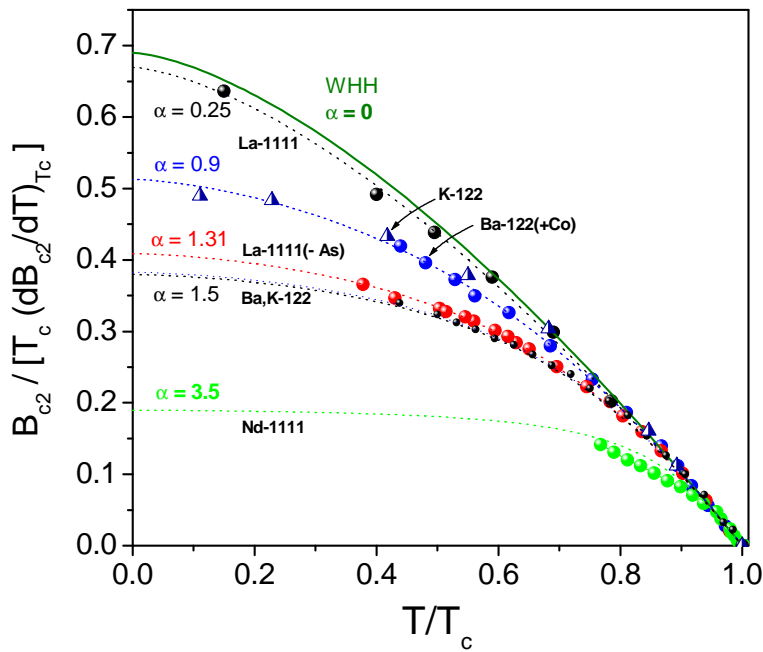}
\caption{Reduced upper critical field 
$B_{c2}/(T_c$[d$B_{c2}$/d$T$]$_{T_c}$) vs $T/T_c$ for the $B_{c2}(T)$
 data shown in figure~11. Dotted lines: WHH model for the indicated 
$\alpha$-values; solid line: WHH model for $\alpha= 0$.} 
\label{f15}
\end{figure}
The paramagnetic pair-breaking of selected compounds is compared in figure~12
where the normalized upper critical field 
$h^* = B_{c2}/ \left[ T_c (dB_{c2}/dT)_{T_c}\right]$ is plotted
against the reduced temperature $t = T/T_c$.
Besides our data for an As deficient La-1111 sample
data from other systems obtained in references 
\cite{yamamoto,altarawneh,jaroszynski} have been included.
The Maki parameter 
was found to increase from $\alpha = $0.25 for 
LaO$_{0.93}$F$_{0.07}$FeAs \cite{kohama-comp} 
over $\alpha = 0.9$ for Ba(Fe,Co)$_2$As$_2$ \cite{yamamoto} and 
$\alpha = 1.31$
 for our As-deficient La-1111 sample up to about 
$\alpha = 3.5$ for NdO$_{0.7}$F$_{0.3}$FeAs \cite{jaroszynski}. 
Except for the last sample, 
the experimental data can be well described by the WHH model using the 
approximation mentioned above. As expected, the deviation of $h^*(t)$ at 
low temperatures from $h^*(t)$ for $\alpha = 0$ increases with $\alpha$
 due to rising 
paramagnetic pair-breaking. The suppression of $h^*(t)$ at low temperature 
is mostly pronounced for NdO$_{0.7}$F$_{0.3}$FeAs. We estimated for 
this compound an upper critical field of $B_{c2}^p(0) \sim$80~T at 
$T = 0$ (see table~1). The importance of paramagnetic effects 
on the $B_{c2}(T)$ data of NdO$_{0.7}$F$_{0.3}$FeAs was also pointed out by 
Jaroszynski {\it et al.} \cite{jaroszynski}. 
They analysed their $B_{c2}(T)$-data within a two-band model for 
dirty-limit superconductors \cite{gurevich}
 and 
estimated $B_{c2}^p(0) \sim 130$~T.  Within this approach,  the kink in 
$B_{c2}(T)$ at about 30~T which is clearly visible for 
$H \parallel a$ \cite{jaroszynski}
could not perfectly described \cite{remgurevich}. However, 
we consider 
this kink as 
indication for the influence of paramagnetic effects on $B_{c2}(T)$.  
	
According to figure~12, the paramagnetic pair-breaking in 
Ba$_{0.55}$K$_{0.45}$Fe$_2$As$_2$ is stronger than in the, 
at first glance, more disordered Ba(Fe,Co)$_2$As$_2$  sample and comparable 
with that in our As-deficient La-1111 sample. This apparent 
discrepancy can be resolved by taking into account the real stoichiometry of 
the investigated Ba$_{0.55}$K$_{0.45}$Fe$_2$As$_2$
 single crystals which has been analysed by wavelength dispersive x-ray 
spectroscopy (WDX) \cite{ni}.
In the single 
crystals which were grown by the flux method using a tin  flux, 
about 5\% As vacancies and $\sim $ 0.66\% Sn were found, the latter 
being most 
likely incorporated on As sites \cite{ni}.
Most probably, the As vacancies are created during the preparation process 
of the single crystal in the tin flux due to the stronger solubility 
of As in Sn than of the other parts of the compound therein.
 Thus, the strong paramagnetic pair-breaking in the 
Ba$_{0.55}$K$_{0.45}$Fe$_2$As$_2$ single crystal seems to be attributed to 
the pronounced As deficiency of these samples which is comparable with that 
in our As-deficient La-1111 sample. 
\begin{table}
\caption{Upper critical field data of selected Fe pnictides.
$B^*_{c2}(0)$ and $B^p_{c2}(0)$ denote the  orbital and the paramagnetically 
limited $B_{c2}$ at $T = 0$, respectively; $\alpha$ is the  fitted Maki 
parameter; $B_p(0)$ denotes the   Pauli limiting field.
}
\vspace{0.3cm}
\hspace{0cm}
\begin{tabular}{c|ccc||cccc}
\hline
Compound & $T_c$   & -$\left(\frac{{\rm d}B_{c2}}{{\rm d}T}\right)\mid_{T_c}$  
& $B_{c2}^*(0)$     &$\alpha$ & $B_p(0)$  	& $B_{c2}^p(0)$	&Reference\\
&(K)&(T/K)&(T)&&(T)&(T)&\\
\hline
LaO$_{0.93}$F$_{0.07}$FeAs           & 25.0 & 4.2& 72
                                     & 0.3 & 305 & 69 
                                                       &\cite{kohama-comp}\\
Ba(Fe$_{0.9}$Co$_{0.1}$)$_2$As$_2$& 21.9 & 4.9&74&0.9&116&
55&\cite{yamamoto}\\
LaO$_{0.9}$F$_{0.1}$FeAs$_{1-\delta}$& 28.6 & 5.4 & 106 
                                     & 1.3 & 114 &63  
                                                       &this work,
                                                        \cite{fuchsprl}\\
(Ba$_{0.55}$K$_{0.45}$)Fe$_2$As$_2$& 32.0 & 6.3 & 138&1.5&130&76
                                                 &\cite{altarawneh}\\
KFe$_2$As$_2$& 2.8 & 3.2 & 6.2&0.9&9.7&4.6
                                                 &\cite{terashima,xtera}\\
NdO$_{0.7}$F$_{0.3}$FeAs&45.6&9.3&293& 3.5&  118&  80   
                                          &\cite{jaroszynski}\\
\hline
\end{tabular}
\label{t1}
\end{table} 
\subsection{Aspects of anisotropy and multiband superconductivity}
In general it should be noticed that on one hand a sizable anisotropy
both for the upper critical fields $B_{c2}$ as well as of the penetration depths
is not surprising in view of the layered structure of the
Fe pnictide superconductors. On the other hand  the different temperature
dependence of the anisotropy for the penetration depth and the upper critical 
field has been claimed to be a great puzzle \cite{prozorov,karpinski}.
Albeit we are
still not able to explain all strange anisotropies, at least several
most striking observations regarding the 
upper critical field with an almost
{\it vanishing},"confluence"-like, anisotropy at low-temperature such as in 
SrFe$_{2-x}$Co$_x$As$_2$ \cite{baily} near 45~T and in 
Ba$_{0.6}$K$_{0.4}$Fe$_2$As$_2$ \cite{yuan} near 55~T or with a strongly 
reduced anisotropy of $\gamma_B=B^{a,b}_{c2}(T)/ B^c_{c2}(T)\approx 1.2-1.5$
 such as in Ba$_{0.55}$K$_{0.45}$Fe$_2$As$_2$ \cite{altarawneh}
(probably with As vacancies)  near 60 T, and in  BaFe$_{2-x}$Co$_x$As$_2$
 \cite{yamamoto} near 45~T can be explained simply by the earlier 
onset of the PLB 
for the 
in-plane component, i.e.\ for $H\parallel$ (a,b). 
Noteworthy, a similar effect with $\gamma_B\approx 1.9$,only,
 has been reported for the one-layer cuprate
superconductor Bi$_2$Sr$_2$CuO$_{6+\delta}$ by 
Vedeneev {\it et al.} \cite{vedeneev}.
We note that the reported deviation between their observed upper critical 
field $B^{ab}_{c2}$=52~T and the theoretical one 
can be removed introducing a moderate strong coupling 
correction $1+\lambda$, with $\lambda=0.625$,
for $B_p$ as considered above.

The striking "confluence"-like behaviour deserves special attention.
In a simple paramagnetic picture like that used here a 
more or less sharp turn to a common 
flattening of $B_{c2}(T)$ for lower $T$ would be expected. 
If in contrary a significant further common increase of 
all $B_{c2}$-components 
at lower $T$ and higher fields 
will be detected in future measurements, a redistribution of 
electrons
between the charge carrier subsystem, the electronic "glue" of 
the pairing
interaction, and the limiting paramagnetic subsystem should be envisaged.
Alternatively, when the superconducting gap becomes comparable with 
the energy of a coupled bosonic mode,
or for a strongly anharmonic lattice system the electron-lattice
interaction and/or polaronic effects might strengthen resulting in a stronger 
coupled superconducting state.

The different anisotropy ratio 
of 
the penetration depths is probably a multiband effect related to "heavy" holes
being mainly responsible for $B_{c2}\propto v^{-2}_F$ and "fast" electrons 
which dominate the 
penetration depths. In fact, adopting for a crude
estimate the averaged Fermi velocities $v_F$
for the hole and electron bands as calculated by Singh and Du \cite{singh}
for LaOFeAS
one has $v^h_{F,ab}=0.81\cdot 10^5$~m/s and  
$v^h_{F,c}=0.34\cdot 10^5$~m/s for the 
hole bands as well as $v^{el}_{F,ab}=2.39\cdot 10^5$~m/s and 
$v^{el}_{F,c}=0.35\cdot 10^5$~m/s for the electron bands. Ignoring
for the sake of simplicity
the interband interaction (possibly important for the high-$T_c$ value),
we estimate the anisotropy ratio of the upper critical fields at very low
temperature in the clean limit 
$\gamma_B(0)\sim v^h_{F,ab}/v^h_{F,c}=2.38$ in accord with about 2 
estimated from torque measurements 
(extrapolated to $T=0$ from $\gamma_B$(34~K)= 5.21 
for NdO$_{0.7}$F$_{0.3}$FeAs) 
in figure~4 of reference \cite{weyeneth}.
Since this estimate is based on the orbital $B_{c2}$ 
ignoring the 
possible PLB,
it should be regarded as an upper bound.
For the analogous penetration depth quantity one estimates
$\gamma_{\lambda_L}=v^{el}_{F,ab}\left(1+\delta^{el}_c \right)/
\left( v^{el}_{F,c}
(1+\delta^{el}_{ab})\right)
=6.82\left(1+\delta^{el}_c\right)/\left(1+\delta^{el}_{ab}\right)$,
where $\delta_i \sim \xi_{0,i}/l_i$ is a parameter which 
measures the effect of disorder on the anisotropic penetration depth 
$\lambda_{L,i}(0)$, $\xi_{0,i}$ is the anisotropic coherence length
 and $l_i$ denotes the corresponding free mean 
path \cite{waelte}. In order to reproduce the large experimental value
of 19 \cite{weyeneth}, significantly anisotropic scattering rates 
 $\delta_z \gg \delta_{ab}$ must be naturally assumed.
Noteworthy, a similar anisotropy and moderate dirtyness has been 
observed for NbSe$_2$ according to its analysis given by 
Bulayevski \cite{bulayevski}. There the in-plane mean free path
exceeds the inter-plane one by a factor of two to four.
A more detailed analysis is hampered by the lacking
information of the experimental partial transverse plasma frequencies. 
If the introduced disorder affects the interband scattering, a slightly
reduced $T_c$ would be expected. However, the suppressing of remnants 
of (fluctuating) SDW antiferromagnetism and a possible additional pairing
attributed to the polarization of charges localized near the As vacancies
\cite{drechsler09} might even overcompensate the former effect.
Even if the disorder will not seriously affect the interband scattering due to 
different symmetry of the states which form the 
electron and the hole FSS  as suggested in
\cite{senga1,senga2}, 
there must be a relevant intraband scattering mechanism which is responsible 
for the strong increase of $B_{c2}(T)$ at high $T$ near $T_c$ and 
relatively low 
external fields (below 30~T).  

At least for highly disordered samples
with strong enough interband scattering
an unconventional scenario is very unlikely.
The limiting Pauli-field $B_p$ can be estimated in a two-band situation by
\begin{equation}
B_p\mbox{[Tesla]}=1.06\Delta_1\mbox{[K]}(1+\lambda_1)^{\varepsilon}\sqrt{\frac{N_1}{N}}
\sqrt{1+\frac{\Delta^2_2N_2(1+\lambda_2)^{2\varepsilon}}{\Delta^2_1N_1(1+\lambda_1)^{2\varepsilon}}} \ ,
\end{equation} 
which generalizes equation~(4). Here and before the occurence of a 
first order transition related to a FFLO-type state
(Fulde-Ferrel-Larkin-Ovchinnikov) at low $T$
has been ignored for the sake of simplicity and the fact that 
such a situation seems 
to be still beyond our available high-field range. Its influence
has been considered in the weak
coupling limit for the case of a two-band
superconductor with very weak interband coupling by Dias \cite{dias}. 

Finally, we note that similarly as in reference \cite{prober}
devoted to the study of highly anisotropic intercalated 
transition metal dichalcogenide layered compounds, the formation of
some microshorts between the superconducting layers cannot be ruled out.
Such microshorts might contribute to the interlayer coupling and to a
somewhat reduction of field penetration in the interlayer spacing for
parallel external fields. This way the anisotropy ratio $\gamma_B$ would
be also reduced. In the context of weakly coupled superconducting layers
the observation of a crossover from 3D to 2D-fluctuations 
with increasing $x$ from underdoped to optimally doped cases 
and a huge slope of $B^{ab}_{c2}(T)$ near $T_c$ up to -11 to -12~T/K 
for SmO$_{1-x}$F$_{x}$FeAs  \cite{palecchi} is noteworthy.
Whereas 3D-fluctuations above $T_c$ have been reported 
in reference \cite{salem-sugui} for   
Ba$_{1-x}$K$_x$Fe$_2$As$_2$ single crystals.

\subsection{Possible origin for the Pauli-limiting behaviour}
According to the classical wisdom  the Pauli susceptibility 
$\chi_{sp}$ is
the central
physical quantity being responsible for the pair-breaking of singlet 
Cooper-pairs \cite{clogston,chandrasekhar,orlando,schossmann,bulayevski}
(see also figure~10~(right) for a schematical view).
Its enhancement causes a lowering of the Pauli limiting field $B_p$.
In fact,  
in our As-deficient sample a significantly  {\it enhanced} 
induced magnetic 
moment has been observed in the normal state above $T_c$
applying weak external magnetic fields \cite{klingeler}. Reducing the 
electron doping 
a similar increase of $\chi(T)$ by a factor of two relatively to optimally doped 
LaO$_{0.9}$F$_{0.1}$FeAs  
with a maximum just at the boundary between the
AFM commensurate SDW and the paramagnetic state has been observed
at $x=0.05$ by Nomura {\it et al.} \cite{nomura}. 
Its suppression for smaller $x$ is a natural consequence of the 
opening of the SDW-gap and the corresponding lost of density of states (DOS) 
at the 
Fermi energy $N(E_F)\propto \chi_{sp}(T=0)$. Suppressing the nesting induced 
competing antiferromagnetism and its 
fluctuations by the As vacancies, the ferromagnetic fluctuations measured
directly by  $\chi_{sp}$ (being the response function for ferromagnetic ordering)
may further increase beyond the value achieved in As stoichiometric samples
at $x=0.05$.
In this context it is noteworthy that the observation of 
ferromagnetic spin fluctuations has been reported by 
Kohama {\it al.} \cite{kohama-fm}. 
This  might explain the missing Pauli-limiting behaviour in stoichiometric 
but 
underdoped samples. Microscopically, the disorder caused by As-vacancies
will affect the total Fe~$3d$ exchange integrals due to a modulation of the 
superexchange admixture involving the As-$4p$ orbitals. 
The theoretical 
difficulty of such a scenario 
consists in its essential physics beyond the mean-field type behaviour
favoured in the present case by the quasi-2D nature of the 
antiferromagnetism. The latter 
manifests itself in the increasing $\chi(T)$ with an expected 
 maximum somewhere above 300~K
for the clean samples \cite{klingeler2} in contrast with an {\it decreasing} 
$\chi(T)$ at the mentioned above high-level for 
our sample  at least for  375~K~$\geq T>T_c$ \cite{klingeler}.
Further theoretical and experimental work is necessary
to make this qualitative scenario more quantitative to be checked in detail.
In this context the observation of very strong FM spin fluctuations
and possibly also of an unconventional 
spin triplet $p$ or $f$ wave
pairing as in superfluid $^3$He or in Sr$_2$RuO$_4$ for  the closely related 
LaOFeP system by Kohama {\it et al.} \cite{kohama-pf} is noteworthy.
An unconventional pairing symmetry {\it different} from that in 
other pnictide
superconductors 
 has been also proposed 
by Fletcher {\it et al.} \cite{fletcher} based on a linear $T$-dependence
of the penetration depth down to 100~mK.
The close vicinity of this remarkable system to a multiple critical point
is also illustrated by the absence of any
 ordering, neither superconducting
nor FM or AFM one
but again by the observation of some
FM spin fluctuations
(being, however, relatively weak,
 if a free electron Landau diamagnetic contribution to the total spin
susceptibility is adopted)
according to reference \cite{mcQueen}.
Here the missing $p$-wave superconductivity might attributed to 
relatively strong disorder reflected by the 
residual resistivity exceeding by an order of magnitude the 
values reported for superconducting samples with $T_c\approx 6$~K
\cite{hamlin,kamihara2}. Without doubt, inspite of its low $T_c$, 
LaOFeP is one of the most challenging pnictide superconductors due to the
vicinity of at least four competing phases. The general 
necessity to improve
the LDA-calculations for such systems
with respect to fluctuations has been pointed out for instance in 
recent references 
\cite{larson,mazin4,singh3}.

Thus, LaOFeP seems to show no AFM ordering due to a less 
pronounced nesting and/or a stronger influence of competing FM 
fluctuations compared with LaOFeAs. In our case of As-deficient
LaFe oxyarsenide
a corresponding 
$p$ or $f$ pairing can be excluded due to our observation of 
a PLB. Microscopically, it is prevented by the strong disorder resulting 
from the As vacancies.    

Alternatively, the suppressed upper critical field at low $T$ and 
high external 
fields might be attributed to some enhanced local exchange field acting on 
the 
superconducting
charge carriers. Such a local field might be caused by an increasing alignment
of localized extrinsic magnetic moments. In the present context localized 
magnetic
moments might occur on a microscopic level just in the vicinity of 
As vacancies as 
shown
by Lee {\it et al.} \cite{lee-alphaFeSe} for the analogous case of 
Se-vacancies 
for the closely related system 
FeSe$_{0.85}$, albeit a Se-vacancy represents probably 
a weaker perturbation
than the As counterpart due its smaller ionic charge of -2 compared 
with -3 for the latter. Remarkably, the DOS at the Fermi-level is enhanced 
introducing vacancies according to reference \cite{lee-alphaFeSe}. Hence, 
even on 
a one-particle level the spin susceptibility $\chi(0)$ will be enhanced.
A similar effect resulting in an enhanced spin
susceptibility $\chi_{sp}(0)$ and an onset of paramagnetic effects already 
above about 2~T, we would suggest also for LiFeAs \cite{tapp}
  in order to explain its  unusual negative curvature of $B_{c2}$ near 
$T_c$ and the extraordinary large slope of $B_{c2}\sim -8$~T/K at a relatively 
low $T_c$-value of 18~K, only.

Returning to our As-deficient sample we note, that 
localized magnetic moments might occur also by a 
possible electronic phase separation.
The formation of a nonsuperconducting paramagnetic phase up to 
a volume fraction as large as 49~at.\%
for Ba$_{1-x}$K$_x$Fe$_2$As$_2$  
and only about 25~at.\% for the superconducting phase has been reported 
by Park {\it et al.} \cite{park-phase-separation} to be compared with
a significantly larger $\left( \stackrel{>}{\sim}40\ \%\right)$
fraction of the superconducting phase in our case.
The simplest possibility for an enhanced local field 
would be given by a contribution from 
a  present
strongly paramagnetic, AFM or FM secondary phases 
coexisting with the 
superconducting main phase. For instance the AFM compound 
Fe$_2$As
or others might be in high fields converted into a highly polarized magnetic
state.
Irrespective of its relation to the suppressed 
upper critical fields, the last two 'extrinsic'
cases might be explain the observation 
of static magnetic moments reported above in 3.2 devoted to muon 
spin relaxation spectroscopy.

Again, further work is necessary to distinguish between these three 
possibilities.
Independently on the true microscopic origin we are confronted with a very 
unusual situation for competing superconductivity and various magnetic 
instabilities or effects
unknown to best of our knowledge for other competing families such as the 
magnetic borocarbides \cite{mueller} and ternary 
compounds \cite{fisher}. Namely, an enhancend superconductivity and 
upper critical fields (enlarged slopes of $B_{c2}$ near $T_c$ 
see also references \cite{singh-disorder,bharathi-disorder}) 
at high temperature und relative low fields below 30 T 
are opposed
by a weakened superconductivity at high fields above 30 T and 
low temperature below 23 K. 

\section{Conclusions and Outlook}

To summarize, we reported a high-field study of  
LaO$_{0.9}$F$_{0.1}$FeAs$_{1-\delta}$ samples with improved 
superconductivity 
near $T_c$. At lower $T$ and very high fields, however, a flattening of 
the $B_{c2}(T)$-curve points to Pauli limiting 
with $B_{c2}(0)\approx$ 63~T to 68~T extrapolated. 
A similar behaviour can be 
deduced 
also from other disordered systems.
In particular, we interpret the flattening of the $B_{c2}(T)$-curve reported 
for
Ba$_{0.55}$K$_{0.45}$Fe$_2$As$_2$ single crystals in 
reference \cite{altarawneh} 
for $H \parallel a,b$ as a strong indication for Pauli limiting. 
Disorder in 
these single crystals is due to about 5~at.\% As vacancies appearing
during their preparation in a Sn flux probably due to the stronger solubility 
of As in Sn than for the other parts of the compound.
In view of the achieved improved d$B_{c2}/$d$T$ 
near $T_c$ and the enhanced $T_c$-value, the introduction of As vacancies or 
of other appropriate 
defects opens new routes for optimising their properties. We are 
confronted with a rather
unusual situation not observed so far to the best of our knowledge: Improved 
superconductivity at high temperature and low fields and somewhat suppressed 
superconductivity at high-fields and low temperature. The first observation 
and 
the enhancement of $T_c$ can be understood within conventional $s$-wave 
superconductivity by enhanced disorder and by a disorder-induced 
suppression of 
nested related remnant antiferromagnetism, respectively.
The relative weakening of the high-field properties strongly suggests an 
enhanced
paramagnetism as the preceding state to a ferromagnetism phase. In other 
words,  
in disordered pnictides superconductivity seems to
compete with at least {\it two} kinds of 
magnetism. Further investigations devoted to a more detailed study of this 
interplay, for instance between the actual As vacancy concentration 
and a 
variation of the electron doping by changing the F and/or the La 
content or introducing also
oxygen vacancies are of considerable basic and technological interest. The 
elucidation of the microscopic reason of 
the observed 
 anomalous high-field properties 
will be obviously helpful for the understanding of the still unknown pairing 
mechanism, too. In particular, the striking
"confluence"-like behaviour of the 
anisotropy near 45 to 60~T in various 122-systems, 
makes a more detailed investigation especially at still higher fields strongly
desirable. 
On the basis of our results for $B_{c2}(T)$
for relatively low magnetic fields and high temperatures, 
and the mentioned above 
increasing hints for $s_{\pm}$ symmetry realized in the clean and 
quasi-clean limits,
two alternative scenarios of opposite disorder influence 
might be suggested:
scenario (i): an impurity driven change of the pairing state
from unconventional $s_{\pm}$ to conventional $s_{++}$ superconductivity
and scenario (ii): a special impurity driven stabilization of the 
unconventional
$s_{\pm}$ state due to the suppression of remnant pair-breaking interband 
scattering.
In scenario (i) the reduction of the clean multiband $T_c$
due to the smearing out of the multi-gap/anisotropic gap structure 
and the unfavorable negative chemical pressure (see figure 1)
must be 
compensated by a strengthened pairing interaction and/or the 
suppression of competing remnant magnetic effects.
Whereas in scenario (ii) the special As-vacancies are assumed to 
scatter predominatly within the individual FSS only and the 
efficiency of an
unfavorable
interband scattering present 
already 
in the non-As-deficient samples is reduced.
For this reason $T_c$ is increased provided the scattering occurs close 
to the unitary limit in contrast to the weaker scattering described 
approximately by the self-consistent Born approximation in case (i).
Due to the present poor understanding of the scattering properties
of As-vacancies, we are unable to decide which scenario
is realized for our system and further experimental studies are necessary. 
Key experiments would be the observation/nonobservation of the spin
resonance mode in inelastic neutron scattering or other signatures
of the $s_{\pm}$ state. In general, a detailed theoretical and experimental 
study of the scattering properties for various relevant impurities is very 
important for properly
understanding the superconductivity in FeAs based compounds.

The Pauli limiting found her suggests that measurements
should be continued  
at least up to 70~T-100~T 
in order to eludicate, whether there is still much 
room for increasing $B_{c2}$ beyond that range. Apparently, the solution of 
this 
problem will affect the evaluation of future high-field applications
of Fe based arsenides and related systems 
in dependence of the requested external field regime.

{\it Note added.} In preparing a revised version of the 
present manuscript to be compared with \cite{fuchsarxiv},
we have learned about an upper critical field study by Terashima
{\it et al.} \cite{terashima} of the low-$T_c$ overdoped 
superconductor KFe$_2$As$_2$  
which also nicely fits our PLB scenario.
For this reason we have re-analyzed their data 
in the same manner as for the other five systems given in table~1
adopting 
$\lambda_{s0}=0$
and included them in 
figure~12, too \cite{xtera}. This less complex
system with $T_c \approx 2.7$ K is of special interest because there
the multiband corrections are almost negligible \cite{rosner} and the large
initial slope is related mainly to the intraband scattering on the dominant 
hole Fermi surface sheets. The observed relatively large
slope is in  accord with our picture of strong 
intraband scattering on hole and electron FSS in the general case. Since 
this single
KFe$_2$As$_2$ crystal has been grown 
also
from a tin-flux we ascribe the scattering 
centers also to As-vacencies. 

The case of strong interband scattering and the corresponding
$s_{++}$-state has been considered by Kuli\'c {\it et al.} \cite{kulic}.

\ack
We kindly acknowledge H~Eschrig, K~Scharnberg,
K-H~M\"uller, R~Klingeler, C~Hess, A~Gurevich, 
I~Mazin, 
O~Dolgov, M~Kuli\'{c}, I~Eremin, and V~Gvozdikov
for discussions. Part of this work has been supported by the EuroMagNET under 
the EU contract RII3-CT-2004-506239 and the SFB 463 (FG, DSL). GB and AE thank 
the DFG for financial support under contracts Be1749/11 and Be1749/12.
AE is also grateful to the IFW Dresden for hospitality.

\newpage
\centerline {SUPPLEMENTARY MATERIAL}
In this supplementary part we collect some transport data which might be helpful
for comparison and characterization with other As-deficient 
 samples to be possibly prepared and investigated by other authors
 in the future.\\

{\bf S1 Hall data}\\
The field dependency of the Hall resistivity $\rho_{xy}$ of the 
As-deficient sample
 shown in figure~13~(left) at $T_c \leq T \leq$~110~K reveals a 
linear
 dependence of $\rho_{xy}$ on the applied magnetic field $H$ and a 
slight temperature
 dependence of the slope $\rho_{xy}/H$. The resulting Hall coefficient 
$R_H =\rho_{xy}/\mu_0H$  
deviates from that known for optimal doped samples,
as shown in figure~13 (right).

\begin{figure}[b]
\includegraphics[width=7.2cm]{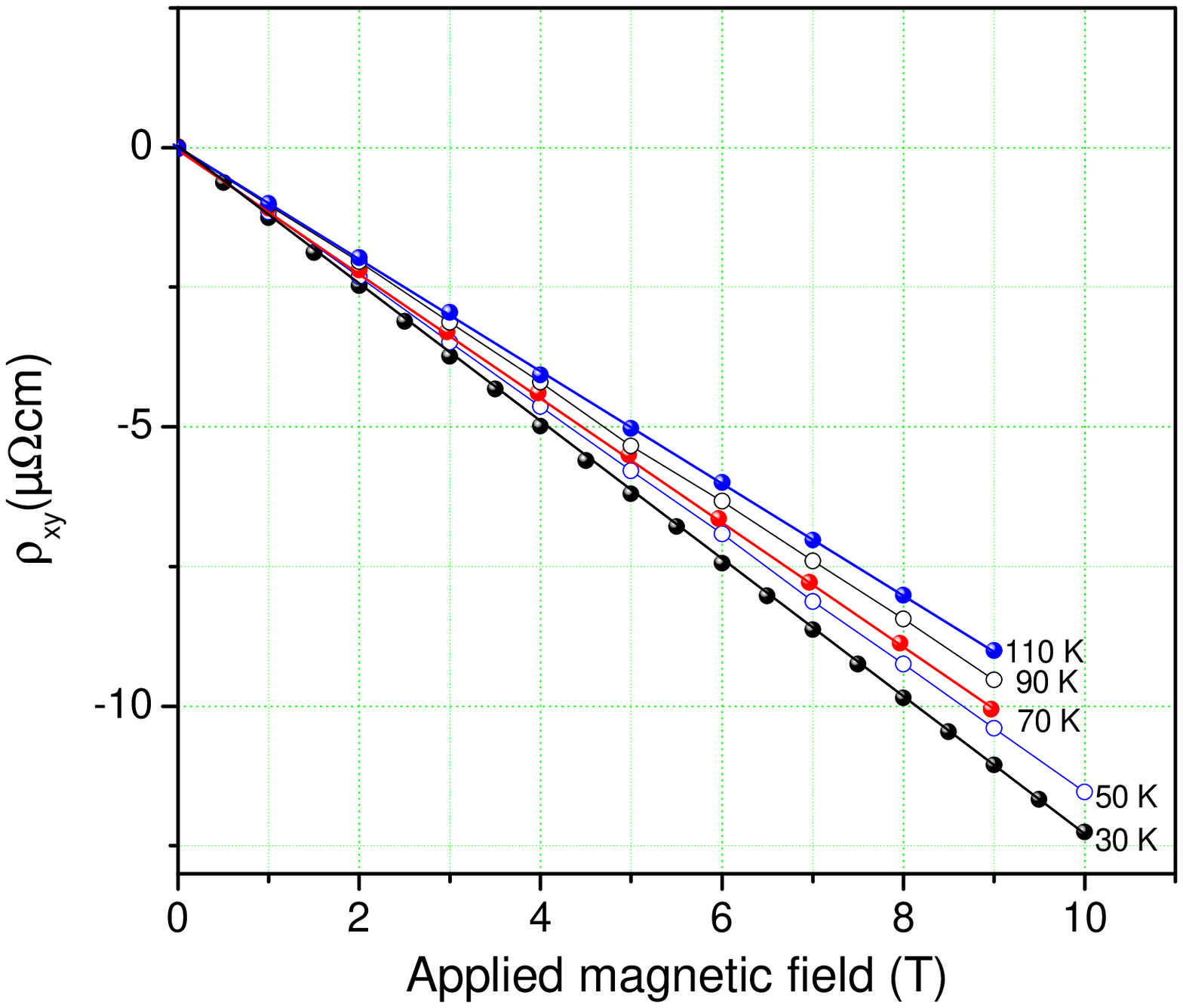}
\hspace{0.5cm}
\includegraphics[width=7.5cm]{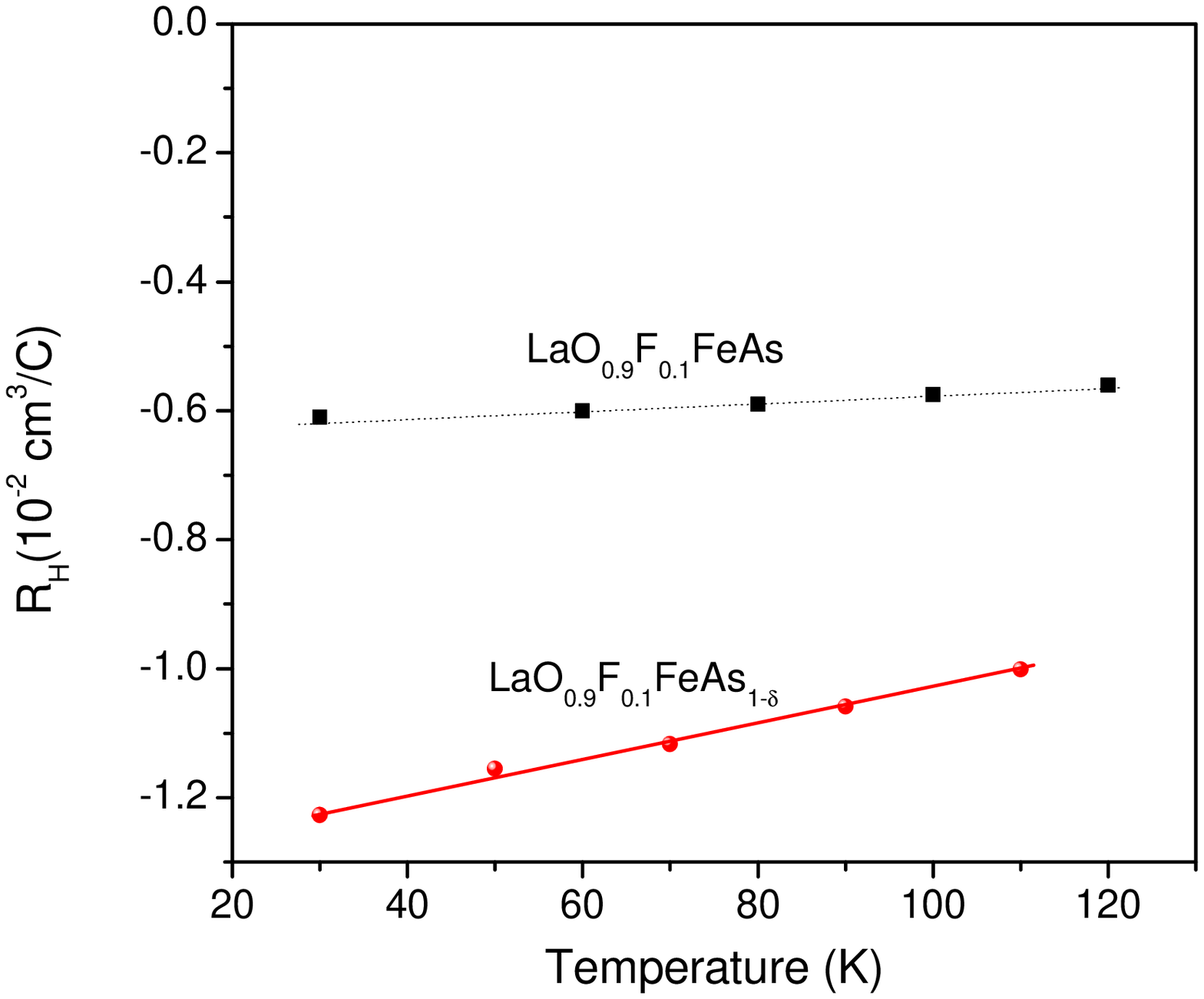}
%
\caption{Field dependence of the Hall resistivity for the 
As-deficient sample (left). Temperature dependence of 
the Hall coefficient (right) for the As-deficient sample 
(\textcolor{red}{$\bullet$}) and 
a clean sample ($\blacksquare$)as taken from 
reference \cite{kohama-hall}.}
\label{f13new}
\end{figure}
From the Hall coefficient, the carrier density $n_H$ has been estimated
within the single band approach both for 
optimally and underdoped LaF$_x$O$_{1-x}$FeAs 
[S1,26,35,38]
using the  relation $n_H = \left( e \mid R_H \mid \right)^{-1}$.
Within this approach
  a charge carrier density of 
$n_H =  0.52\cdot 10^{21}$cm$^{-3}$ is estimated at $T = 30$~K for the 
As-deficient sample, which is 
comparable to $n_H =  0.55\cdot 10^{21}$cm$^{-3}$ reported for 
underdoped LaF$_x$O$_{1-x}$FeAs with x = 0.05 \cite{kohama-hall},
but only half as large as $n_H \approx 1.0\cdot 10^{21}$cm$^{-3}$
[S1,26,35]
reported for optimally doped 
LaF$_x$O$_{1-x}$FeAs with $x = 0.1$. The reduced carrier density of the 
As-deficient sample points to a reduced doping level of $x = 0.05$ of 
this sample instead of $x = 0.1$ expected from its F content. This is 
consistent 
with the observed enhancement of the lattice parameters of this sample which 
are close to those reported for $x = 0.5$.\\

{\bf S2 Scaling analyzis of the resistivity}\\ 
In order to get more insight in relevant scattering mechanisms
we performed a scaling analysis of the resistivity $\rho(T)$ data from
$T_c$ up to 300~K. Such an analysis is also
helpful to classify our sample
with respect to other underdoped and overdoped
samples of the same La-1111 family. For that purpose we employed
the following expression for the resistivity $\rho(T)$ which has been used 
for cuprate HTSC [S2,S3,S4,S5]
\vspace{2mm} 
$$\hspace{3cm}\rho =\rho_0+CT\exp \left(-\frac{\Delta}{T} \right), \hspace{6.5cm} (S2,1)$$
where $\rho_0$  is the residual resistivity  
and $\Delta$ is a  characteristic energy
determined from the  nonlinear part of $\rho (T)$. 
Expression (S2,1) fits nicely our data (see figure 14 (left).
\begin{figure}[b]
\includegraphics[width=7.9cm,angle=0]{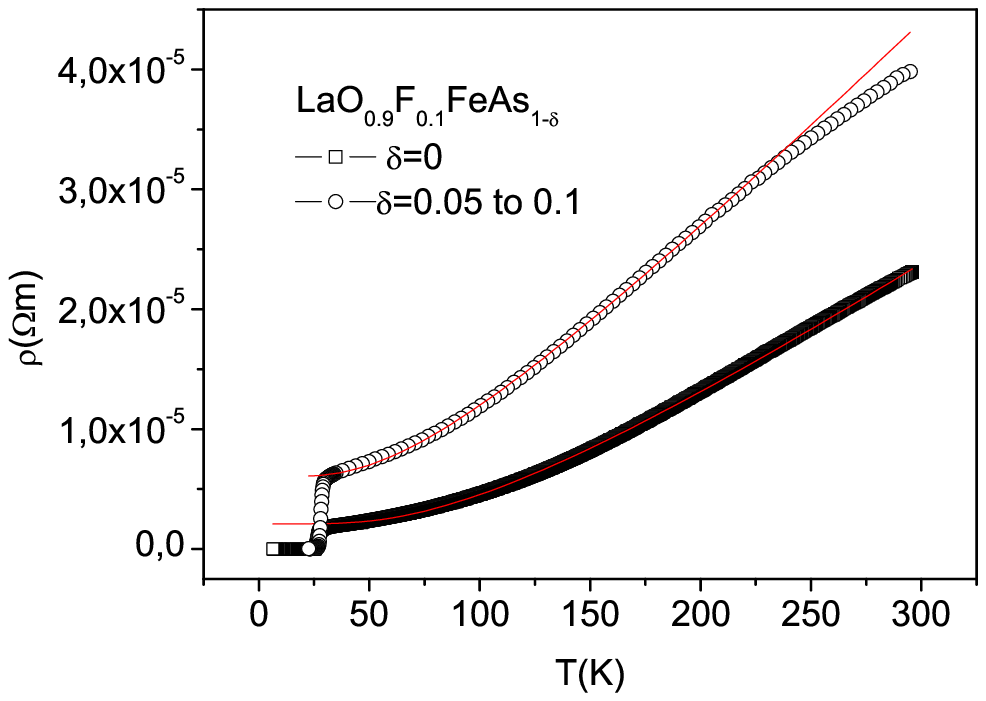}
\includegraphics[width=7.3cm,angle=0]{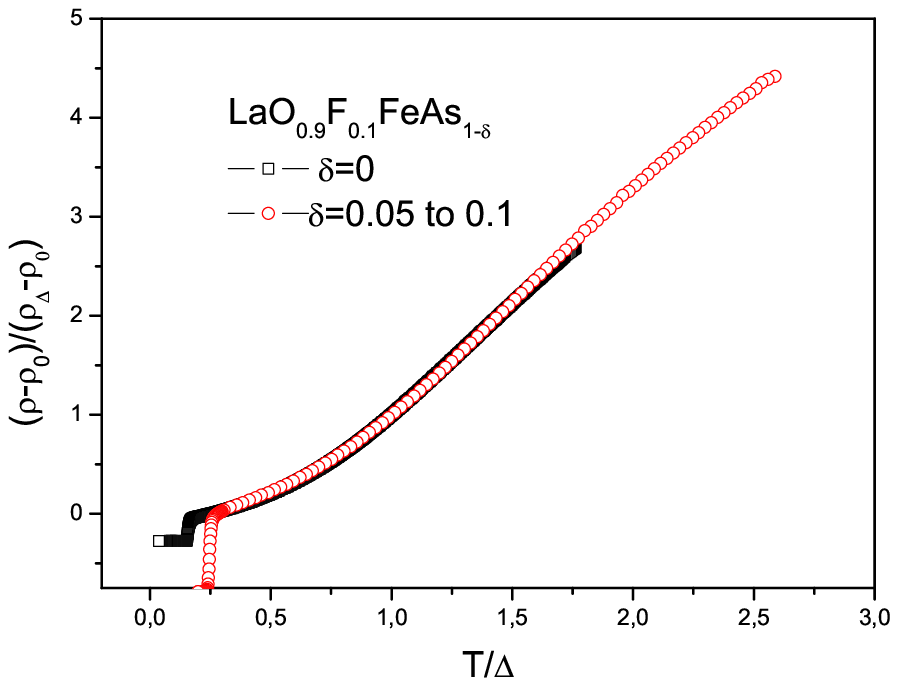}
\label{scalfig}
\caption{Temperature dependence of the resistivity $\rho$ for 
an As-deficient and a nondeficient clean reference sample 
(left). The solid line represents a fit 
using equation~(S2,1). The same as in the left part with eliminated
residual resistivity $\rho_0$ and normalized in an appropriate way (see text)
(right).
}
\end{figure}
The obtained fit parameters $\rho_0$, $C$, and $\Delta$ are shown
in  table~2.
\begin{table}
\caption{Scaling analysis of the resistivity $\rho(T)$ according to
 equation (S2,1), where $\rho_0$ denotes the residual resistivity 
and $\Delta$ is
a pseudo-gap-like quantity.} 

\begin{tabular*}{\textwidth}{@{}l*{16}{@{\extracolsep{0pt plus
12pt}}c}}
\br
Samples&	$\rho_0$&$\rho$(300~K)/$\rho_0$& $\Delta$&C\\
 &($\mu\Omega$cm)&&(K)&($\mu \Omega $cm/K)\\
\hline
LaO$_{0.9}$F$_{0.1}$FeAs$_{1-\delta}$ &
210.0&10.95&164&12.5\\
($\delta=0$)&&&&&&\\
\hline
LaO$_{0.9}$F$_{0.1}$FeAs$_{1-\delta}$  &
605.6&	6.6	&114&	18.5\\
($\delta \approx 0.05$ to $0.1$)&&&\\
\hline
\br
\end{tabular*}
\end{table}
Our values of $\Delta$  are about 14 and 10 meV at $\delta=0$ 
and $\delta=$ 0.05 to 0.1, respectively. Noteworthy, the former value 
is in accord with the so-called pseudo gap of 15-20 meV reported by 
Sato {\it et al.} [S6] 
and Garcia {\it et al.} [S7] 
These estimates are based on high-resolution photoemission spectroscopy
data for 
LaO$_{0.93}$F$_{0.07}$FeAs and LaO$_{0.9}$F$_{0.1}$FeAs, respectively. 
Both curves depicted in figure~15~(left) can be lumped into a single 
curve 
as illustrated in figure~14~(right), if one plots
$(\rho -\rho_0)$/($\rho (\Delta)-\rho_0 )$ against $T/\Delta$,
where $\rho(\Delta)$ is the resistivity at $T=\Delta$.
The resulting curve is approximately 
linear with $T$ for $1.5  \geq T/\Delta \geq 0.8$ 
(see figure~14~(right)). 
At lower temperatures 
the $\rho (T)$-curves deviate from linearity and a superlinear 
$\rho (T)$-behavior sets in. A similar behaviour has been reported 
for underdoped 
YBa$_2$Cu$_3$O$_{7-y}$ [S2,S3,S4]
The existence of a universal metallic $\rho (T)$-curve 
points to a single mechanism which dominates the 
scattering of 
the charge carriers in these materials.
A more detailed scaling analysis will be given elsewhere [S8].\\
 
\noindent
\hspace{0.0mm} [S1] Jaroszynski J, Riggs SC, Hunte F {\it et al.} 2008 \PR B {\bf 78} 064511\\
\vspace{0cm}
\noindent
[S2] Moshchalkov VV, Vanacken J and Trappeniers L
         2001 \PR B  {\bf 64} 214504\\
\vspace{0cm}
\noindent
[S3] Vanacken J 2001 {\it Physica} B {\bf 294-295} 347\\
\vspace{0cm}
\noindent
[S4] Vanacken J, Trappeniers L, Wagner P {\it et al.}
2001 \PR B {\bf 64} 184425\\
\vspace{0cm}
\noindent
[S5] Luo HG, Sum YH, and Xiang T 2008 \PR B {\bf 77} 014529\\ 
\vspace{0cm}
\noindent
[S6] Sato T, Souma S, Nakayama K
{\it et al.} 
2008 {\it J.\ Phys.\ Soc.\ Jap.\ } {\bf 77} 063708\\
\vspace{0cm} 
\noindent
[S7] Garcia DR, Jozwiak C, Hwang CG, 
2008 {\it arXiv:}0810.3034\\
\vspace{0cm}
\noindent
[S8] Arushanov E {\it et al.} to be published

\begin{thebibliography}{99}
\bibitem{rems}In the clean interband limit 
with repulsive interband interaction
the $s_{\pm}$ state is always realized for any attractive intraband interaction.
The proposed strong coupling description of the superconductivity necessary
to reproduce a $T_c \approx$ 26~K with
$\lambda_{12}=-\lambda_{21}= 2$ (!),
$\lambda_{11}=\lambda_{22}=0.5$
and a spin fluctuation frequency 
of about
$\hbar \omega_{sp}=25$~meV proposed in reference 
\cite{dolgov09}
is inconsistent with the weak coupling approximation usually employed for the
desription of the pure AFM instability (see 
e.g.\ references \cite{korshunov,chubukov}).
To the best of our knowledge, at present there is no strong coupling theory 
which
treats magnetism and superconductivity on equal footing even in the
simplest case of the nonrealistic clean limit.
In addition, the calculated penetration depth is in conflict with available
experimental muon spin rotation resonance \cite{luetkens} and optical data
\cite{drechsler,drechsler09} for LaO$_{0.9}$F$_{0.1}$FeAs
\bibitem{xtera}In their analysis of the PLB for the 
in-plane $B_{c2}(T)$ of a KFe$_2$As$_2$ single crystal 
the authors used a slightly different approach as we did. They 
adopted a BCS-like value (i.e.\ $\lambda=0$) for the Pauli-limiting 
field $B_{p}(0)=1.84T_c$
thereby ignoring any 
strong coupling corrections (compare equation (5)). Notice the slightly
different values for the orbital field $B^*_{c2}(0)$ of 7.4 (8.4)~T and 6.2 T
resulting from different slopes of -3.8~T/K and -3.2~T/K 
in their and 
our analysis, respectively. Finally, using their 
original 
data we estimate
$\lambda \sim 0.7$, i.e.\ strong coupling corrections cannot be ignored 
even for this low-$T_c$ superconductor.
\end{thebibliography}
\end{document}